\documentclass[twocolumn,a4paper,nofootinbib]{revtex4-1}

\usepackage[utf8]{inputenx}
\usepackage{graphicx,lmodern,microtype}
\usepackage[T1]{fontenc}
\usepackage[hidelinks]{hyperref}

\newcommand{\eqref}[1]{Eq.~\ref{#1}}
\newcommand{\degree}{\ensuremath{^\circ}}

\begin{document}

\title[Langevin power curve analysis (published in {\it Wind
  Energy})]
{Langevin power curve analysis for numerical WEC models\\
  with new insights on high frequency power performance}

\author{Tanja A.\ Mücke}
\author{Matthias Wächter} 
\author{Patrick Milan} 
\author{Joachim Peinke}

\affiliation{ForWind -- Center for Wind Energy Research, Institute of
 Physics, University of Oldenburg, Germany
} 

\collaboration{
  Published in: Wind Energy (2014), doi: 10.1002/we.1799
}

\begin{abstract}

\noindent
Based on the Langevin equation it has been proposed to obtain power curves for wind turbines from high frequency data of wind speed measurements $u(t)$ and power output $P(t)$. 
The two parts of the Langevin approach, power curve and drift field, give a comprehensive description of the conversion dynamic over the whole operating range of the wind turbine. 
The method deals with high frequent data instead of 10\,min means. 
It is therefore possible to gain a reliable power curve already from a small amount of data per wind speed. 
Furthermore, the method is able to visualize multiple fixed points, which is e.g. characteristic for the transition from partial to full load or in case the conversion process deviates from the standard procedures. \\
In order to gain a deeper knowledge it is essential that the method works not only for measured data but also for numerical wind turbine models and synthetic wind fields. 
Here, we characterize the dynamics of a detailed numerical wind turbine model  and calculate the Langevin power curve for different data samplings.
We show, how to get reliable results from synthetic data and verify the  applicability of the method for field measurements with ultra-sonic, cup and Lidar measurements. 
The independence of the fixed points on site specific turbulence effects is also confirmed with the numerical model.
Furthermore, we demonstrate the potential of the Langevin approach to detect failures in the conversion process and thus show the potential of the Langevin approach for a condition monitoring system.
\end{abstract}

\maketitle


\section{Introduction}

Power curves describe the relation between the inflowing wind field $u$ and the electrical power output $P$ of a wind energy converter~(WEC).
Although the current industry standard International Electrotechnical Commission~(IEC) 61400-12-1~\cite{IEC-12} defines a unified procedure,  the so-called IEC power curve, it is still a challenging problem to predict the average power output of a WEC. 
One important difficulty is caused by the non linear relation between $u$ and $P$.  
$P(\langle u \rangle) \not = \langle P(u) \rangle $ holds for any kind of non linear response dynamics, 
particularly for non linear functions between power $P$ and wind speed $u$, like $P(u) \propto u^3$ (see also reference~\cite{Gottschall2008}).  
Here $\langle ... \rangle$ denotes the average. 
The inequality is responsible for the transformation of symmetric wind speed fluctuations to asymmetric power fluctuations. 
A linear averaging procedure like the IEC power curve $P_{IEC}$ will lead to problems if the magnitude of fluctuations is not stationary, as it is often the case for turbulent wind conditions. 
In previous work~\cite{Boettcher2007, Gottschall2008, Milan2009}, it is shown that systematic deviations from the real power curve occur by following the IEC recommendations.
For a simple relaxation model, the dependency of $P_{IEC}$ on the turbulence intensity of the inflow is shown. 
The deviations due to the above mentioned inequality become more pronounced for higher turbulence, especially near rated wind speed.  
The latter effect is well known in the wind energy industry~\cite{gasch}.
Thus, the measured IEC power curve is related to the turbulence distribution at the test site, and the power curve measurements of identical turbines at different locations will lead to different results.
In order to decrease the uncertainty related to site-specific effects, a conversion of IEC power curves from one
test site to another is necessary; see also reference~\cite{Kaiser03, Langreder04}.

A few years ago, the alternative Langevin approach has been proposed to obtain power curves for 
WECs directly from high frequent data of power output and wind speed measurements~\cite{Anahua2007,Gottschall2008}.
This new model is based on the intrinsic short-time dynamical response of a WEC to the high frequency wind fluctuations and describes the actual dynamics of the power output on small time scales. 
The dynamical behaviour of the WEC is quantified by a drift field and the corresponding Langevin power curve (LPC), which consists of the fixed points of the dynamical system.
The strength of the new approach is based on the fact that the characterization relates the wind speed and power output in a direct way, regardless of the intermediate steps of the extraction process.
The complex statistical characteristics of the wind like turbulence, non-stationarity and its higher-order 
statistical features like intermittency  are also included in the model.
A special feature of the concept is that the LPC could also display multiple fixed points in case, for a given wind speed, the WEC control system aims for different power levels, depending on the specific situation~\cite{Milan2011}.
With this method, it should also be possible to describe the power performance of a WEC independent of the turbulence characteristics of the inflowing wind field. 
The last two points demonstrate clearly that the obtained power characteristic applying the dynamical method is of another type than the standard IEC power curve.
The IEC power curve specifies mean values and can (amongst others) be used to describe the annual energy production, which is calculated by combining the corresponding IEC power curve with the Weibull distribution of wind speeds~\cite{burton}.  
In contrast, the LPC is based on fixed points of the actual power dynamics~\cite{Gottschall2008a}.
The LPC can give additional information about the operating condition of the wind turbine, and it is even possible to use it to model the high frequency power delivery of a turbine~\cite{waechter_JoT}.
However, the IEC power curve is a well accepted method to determine, e.g. the annual power production.

Since the introduction of the LPC, the method was used in several publications to analyse recorded operational data of several WECs.
Anahua~{\it et al.}~\cite{Anahua2007} were able to estimate LPCs from measured wind speed data and power output at two different sites. 
The wind speed measurements were taken with a cup anemometer in flat terrain and with an ultrasonic anemometer in complex terrain. 
As a main result, they showed that independently of the measurement device and terrain type, the characteristic 
of the wind turbine power performance was properly reconstructed.
Next to cup and ultra sonic anemometers, also light detection and ranging~(LIDAR) measurements were investigated.
Within the research project ''Research at Alpha Ventus (RAVE)''  W\"achter~{\it et al.} have shown a successful application in power performance characterization of a 5\,MW offshore WEC from power measurements related to LIDAR wind measurements of the Leosphere WindCube~\cite{waechter_dewek}. 
The derived power curves were in very good correspondence with those derived from the cup and ultrasonic anemometer wind measurements used in the same  campaign. 
In the following, Gottschall~{\it et al.}~\cite{Gottschall2008} showed how to improve the method of estimating LPCs. 
The results prove that the dynamical approach grasps the actual conversion dynamics of a wind turbine and enables to gain most accurate results for the power curve independent of site-specific influences. 
A further promising application of the LPC was proposed by Milan~{\it et al.}~\cite{Milan2010, Milan2009}, displaying with a simple numerical model that the method in general is able to monitor the conversion process of a WEC. 
 
In the aforementioned publications, it has been shown how promising the new method is for the application in wind energy. 
Nevertheless, the previous findings are based solely on selected measurements where different recording devices, like LIDAR or a a cup-anemometer on a metmast have been used. 
Until now, it is not fully clarified which effect these wind measurements will have on the LPC.
In order to clarify these effects on the LPC method,  we use synthetic wind speed and power output data, obtained from a numerical WEC model.
In particular, we estimate LPCs from modelled data and show what kind of data is necessary for a successful calculation of the fixed points of the dynamical system.
We will show that the construction of the LPC in principle depends on the sampling rate and the number of signals that fill in the power bins, which in its turn depends on the turbulence intensity and, therefore, the site.
Additionally, we revise the conclusions drawn from measured data.
We prove the influence of different measurement devices, turbulence characteristics and occurring failures during operation on the estimated LPC.  
In order to simulate the power output during normal operation, we use a detailed numerical wind turbine model with a turbulent inflowing wind field.
As a numerical WEC model, the simulation code FAST~\cite{fast} is used, which is also frequently used in the wind industry and provides realistic power output data taking into account detailed properties for specific WECs.

The paper is organized as follows. 
We start with a description of the numerical data used and a short summary about the theoretical background of the LPC.
In section~\ref{pc}, we specify in detail how to estimate a reliable LPC for a numerical WEC model.
The procedure is presented step by step while we show the influence of different sampling rates on the LPC.
Significant LPCs are estimated later on for three different data samplings of the synthetic inflow, which can be related to field measurements with  an ultrasonic anemometer, cup anemometer or LIDAR. 
Additionally, the power output is modelled for two different turbulence intensities of the inflowing wind field.
We can show that the estimated LPC is independent of the measurement device as well as from the site-specific turbulence intensity.
Finally, the ability of the method to detect failures is quantified for two simulated failure scenarios.
It becomes obvious that abnormal operating conditions of the WEC can  be detected easily as they alter the drift field and the fixed points of the LPC.




\section{Numerical data}\label{model}
The simulations of the electrical power output are performed with the program FAST\footnote{Please note that FAST produces a time shift in the data during simulation. The time index of the inflowing wind field and the power output are shifted against each other with the time step (0.5*grid width [m]) /(mean hub height wind speed [m/s]). Without the correction of the time index the estimates of the Langevin power are inaccurate.} designed by National Renewable Energy Laboratory (NREL)~\cite{fast} in connection with the WindPact 1.5\,MW (WP\,1.5) wind turbine model. 
FAST simulates the loads of a wind turbine with a combined modal and multibody dynamics formulation.
The model relates rigid bodies (e.g. nacelle, gears, hub) and flexible bodies (tower, blades and drive shaft) through degrees of freedom.
The flexible elements are modelled by using a linear modal representation~\cite{fastguide}.
The virtual WP\,1.5 wind turbine is a variable-speed pitch-regulated wind turbine and has a rotor diameter of 70\,m~\cite{WP15}.
The turbine's hub height is 84\,m, and rated wind speed is obtained at 11.2\,m/s.
For turbulent wind field simulations, FAST requires a synthetic inflow that covers at least the whole rotor area of the wind turbine.
Therefore, the synthetic wind fields are simulated over a grid of 100$\times$100$m^2$ with nine equally distributed points in 
vertical and horizontal direction around the hub height of 84\,m.
This results in 81 grid points with a distance of 12.5\,m, respectively.
We use TurbSim~\cite{turbsim} to generate such a two-dimensional grid applying the turbulence model of Kaimal~\cite{kaimal}.
In this way, a three-component wind is generated at each grid point that march past the turbine at a given mean wind speed.
For the purpose of keeping our analysis quite generic, the synthetic wind fields are generated without a wind shear. 
Measured 10\,min wind profiles are subject to strong fluctuations and depend highly on the atmospheric and site conditions.  
Therefore, the influence of the wind shear on the LPC is part of a another research project and is neglected in our simulations. 
Additionally, we do not include time delays in our simulations as they are unavoidable during a field measurement with an installed metmast in front of the turbine. 

%
\begin{table*} 
  \begin{center}
    \caption[Summary of the numerical data used for the
    analysis]{\label{tab_seeds} Summary of the numerical data used for
      the analysis. The synthetic wind fields are obtained for wind
      speeds from 5 to 15\,m/s in 0.5\,m/s steps and have a frequency
      of 10\,Hz.}
    \begin{tabular}{|l|c|c|c|c|} 
      \hline 
      {\bf Name} & {\bf $I$} & {\bf Mean wind speeds}& {\bf Seeds per wind speed} & {\bf Database} \\  \hline 
      {\it data5} & 0.05 & 5-15\,m/s in 0.5\,m/s steps & 30 & 105\,h \\ 
      {\it data15} & 0.15 & 5-15\,m/s in 0.5\,m/s steps & 60 & 210\,h \\
      \hline 
    \end{tabular}
  \end{center}
\end{table*}

Two cases with different turbulence intensities, $I$, are investigated in our study. 
We use $I$=0.05 representing offshore sites and $I$=0.15 for moderate terrain.  
In both cases, named {\it data5} and {\it data15}, the turbulent wind fields are generated with mean wind speeds at hub height between 5 and 15\,m/s.
Beginning at 5\,m/s the mean wind speed increases in 0.5\,m/s steps for each generation.
In this way, the whole range from partial to full load operation of the wind turbine is covered. 
For all  those 21 different mean wind speeds, a certain number of 10\,min wind fields is needed each with a different initialization (seed) of the TurbSim random number generator to ensure a broad range of statistical variation.
Thus, we generate 30 time series per wind speed for a turbulence intensity of 5\% and 60 time series for a turbulence intensity of 15\%.
The time duration of each analysed load calculation should be 10\,min, as it is common in the wind industry.
In order to remove transient effects in the simulations, all generated wind time series last 700\,s.
Only the last 10\,min of the calculated data are taken into account for the power curve estimation.
A summary of the numerical data used for the analysis is given in Table~\ref{tab_seeds}.




\section{The Langevin Approach}\label{Langevin}
The Langevin approach models the short-time dynamics of the wind power conversion as a relaxation process, driven by the turbulent wind.
The general assumption for the method is that we can decompose the fluctuating wind turbine power
output into two functions:
(i) the relaxation, which describes the deterministic dynamic response of the wind turbine to its desired operation state, and
(ii) the stochastic force (noise), which is an intrinsic feature of the system of wind power conversion.
In this way, the time evolution of the power output $P$ can be described with the Langevin equation as 
\begin{equation}\label{langevin}
 \frac{d}{dt}P(t)=D^{(1)}(P;u)+\sqrt{D^{(2)}(P;u)} \cdot \Gamma(t) \quad .
\end{equation}
The terms $D^{(1)} (P;u)$ and $D^{(2)} (P;u)$ are called the drift and diffusion coefficients, and they describe the deterministic relaxation and stochastic (noise) temporal evolution, respectively. 
With $( \ldots ;u)$, we denote the conditioning of the process on fixed wind speed values $u$.
The characteristics of the wind turbine's power performance are given by the fixed points (steady states) from the deterministic dynamic relaxation, obtained from the condition $D^{(1)} (P;u) = 0$.
The term $\sqrt{D^{(2)}(P;u)}$ describes the amplitude of the dynamical noise, and $\Gamma (t)$ is taken as independent $\delta$-correlated Gaussian white noise with zero mean.
As initially proposed by Siegert~{\it et al.}~\cite{siegert1998}, the drift and diffusion coefficients can be obtained pointwisely in the $u,P$ phase space according to
\begin{equation} \label{eq:D1}
  D_i^{(n)}(P;u)=\frac{1}{n!}\lim_{\tau \to 0} \frac{1}{\tau} M_{ij}^{(n)}(P,\tau ; u)
\end{equation}
with the first (n=1) and second (n=2) conditional moment $M_{ij}^{(n)}$, respectively, given by 
\begin{equation}\label{eq:M1}
 M_{ij}^{(n)}(P,\tau;u)=\langle [P(t+\tau)-P(t)]^n \mid P(t)=P_j, U(t)=u_i \rangle \quad.
\end{equation}
In order to estimate the coefficients $D_i^{(n)}(P)$, the values of the conditional moments are fitted for each wind speed and power bin by a linear regression over a specific range of values $\tau \in [\tau_1, \tau_N]$. 
More precisely, the path integral of the corresponding Fokker-Planck equation gives the $\tau$-evolution of $M_{ij}^{(n)}$ and allows the reconstruction of the $D_i^{(n)}(P)$ coefficients, for which linear dependency is expected for small $\tau$. 
It has been shown that a linear approximation of the conditional moments~$M^{(1)}$ is exact for sufficient high sampling rates~\cite{Friedrich2002,Gottschall2008b}, and the larger the value of $\tau$, the larger becomes the deviation between estimate and intrinsic function for $D^{(1)}$.
The uncertainty of the conditional moments $\sigma [M^{(1)}]$ depends on the amount of statistical events $N$ contributing to each estimation and is given by 
\begin{equation}\label{errorM}
\sigma [M^{(n)}(P,\tau ;u)]=\frac{\sqrt{\langle \mid P(t+\tau)-P(t) - \langle P(t) \rangle \mid ^{(n)} \rangle}}{\sqrt{N}} \quad .
\end{equation}
The corresponding errors for the drift functions are calculated according to Kleinhans and Friedrich~\cite{kleinhans_error} with
\begin{equation}\label{error}
\sigma [D^{(1)}(P;u)]=\sqrt{\frac{1}{\tau}\frac{D^{(2)}(P;u)}{N} -  \frac{[D^{(1)}(P;u)]^2}{N}} 
\end{equation}
and for $D^{(2)}$ in a similar way.
All errors are defined for a finite time increment $\tau>$0 and a finite number of data points $N$ in the bin.

This approach to analyse the data in the framework of reconstructed Langevin equation may be seen in the context of many other methods for separating deterministic and noise part; see reference~\cite{Kantz2003}.
The advantage of the Langevin approach is that the method is directly derived from rigorous mathematical features for general nonlinear stochastic processes~\cite{FriedrichReport} and its comparable fast convergences~\cite{Racca2005}.
 The different presentations like Ito and Stratonovich are related by transformations; see reference~\cite{Risken1984}.
 In this work, the Ito picture is used. 
 Corrections to the  linear approximation of $M_{ij}^{(n)}$ can be given explicitly~\cite{Gottschall2008b, Risken1984} as well as dealing with non ideal noise~\cite{rinn2013}. 
 Siefert~{\it et al.}~\cite{siefert04} have shown that the deterministic part $D^{(1)}$ can be estimated correctly even for highly correlated noise, which  is important for our work here, as we are interested in the  $D^{(1)}$, i.e. the response-reaction of the WEC to changes of the wind speed.
 The stable fixed points of the power conversion process ($D^{(1)}(P;u)=0$ with negative slope) are then considered to be the LPC.
 More detailed information on the stochastic method can be found in the study of Friedrich~{\it et al.}~\cite{FriedrichReport}. 
 Results on LPCs are given in previous work~\cite{Anahua2007, Gottschall2007, Gottschall2008,  Gottschall2008b,  Gottschall2008a}.




\section{Langevin power curves from numerical data}\label{pc}
This section deals with the question how to get a reliable LPC from the numerical WP\,1.5 wind turbine model. 
Firstly, we will address the impact of the sampling rate on the LPC.
The calculation of reliable conditional moments $M_{ij}^{(1)}$ and therewith the drift coefficients $D^{(1)}$ is of crucial significance for the estimation of the fixed points. 
Thus, we will show in detail the influence of the sampling rate on these both quantities.
In order to present also the procedure of how to estimate the fixed points, the results are presented in a step-by-step process.
Secondly, we will investigate the LPC for the special case of spatially averaged data,  which are gained by, e.g. LIDAR measurements.

The power output is calculated from the previously mentioned turbulent inflowing wind fields acting on the whole rotor plane of the wind turbine.
According to the IEC standard~\cite{IEC-12}, the power curve is based solely on the $u$ component of the wind field at hub height representing a conventional one-point measurement.
Ultrasonic and cup anemometer are characterized by different time resolutions of the local wind fluctuations. 
In contrast to this, LIDAR measurements usually average the inflowing wind field over a horizontal layer in front of the turbine, which will be the topic in section~\ref{lidar}.
In several publications, it has been shown that the LPC can be estimated with these different measuring methods~\cite{Anahua2007, waechter_dewek, Milan2010,Gottschall2008a}, but it is still an open question if there are deviations due to the different methods. With our synthetic model approach, it is now possible to perform a quantitative comparison between these methods.

According to equation~\eqref{eq:M1}, the Langevin method is local in phase space (i.e. in the $u-P$ space), and thus, a binning has to be defined. 
In analogy to the IEC standard procedure~\cite{IEC-12}, the wind speed data sets are here divided into bins $u_i$ of 0.5\,m/s. 
Additionally, a sufficient amount of power bins $P_j$ have to be chosen. 
Based on our experience, the amount of power bins should be at least 20 for each wind speed bin to cover the power data within this wind speed bin and to resolve the fixed points correctly.
In the following, the resolution of the power bins is set to 40\,kW in the case of 5\% turbulence  intensity, which corresponds to approximately 40 equally distributed power bins for one wind speed. 
In the case of higher turbulence intensity, a lower resolution of $\approx$~20 power bins, corresponding to 60\,kW, is set as the data are more spread because of the higher fluctuations in wind inflow and power output. 
Please note that no data averaging is applied for the described binning.
Here, the binning can be understood as some kind of grid to sort the instantaneous high frequent wind speed and power data.

\subsection{Impact of the sampling rate }\label{freq}

\begin{figure*} 
  \unitlength 1.2mm 
  \begin{picture}(0,0)
    \put(-4,15){(a)}
    \put(71,15){(b)}
    \put(-4,-15){(c)}
    \put(71,-15){(d)}
  \end{picture}%
  \centering{
    \begin{minipage}{0.95\textwidth}
      \centering{
        \includegraphics[width=0.45\textwidth]{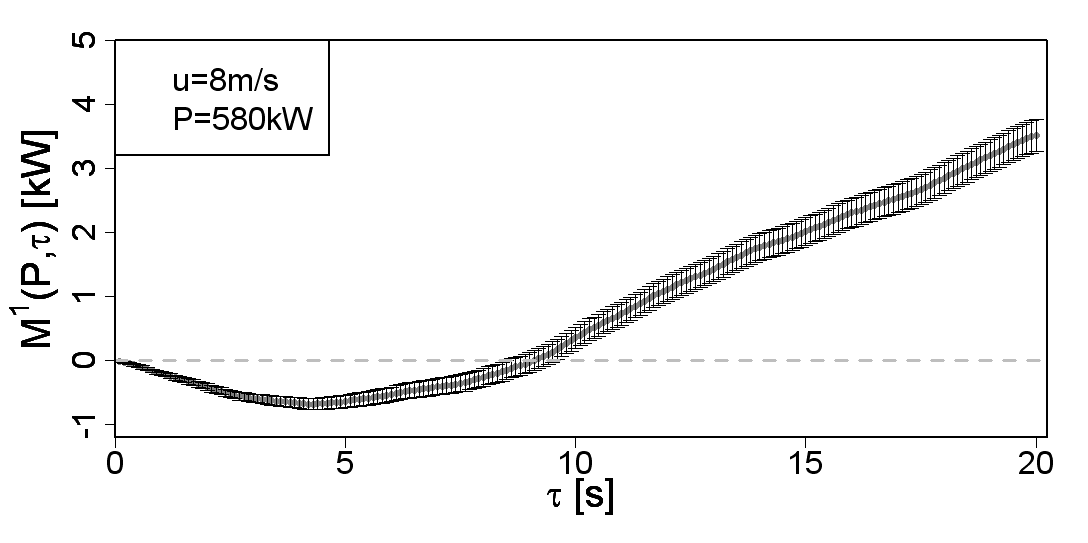}
        \hspace{0.05\textwidth}
        \includegraphics[width=0.45\textwidth]{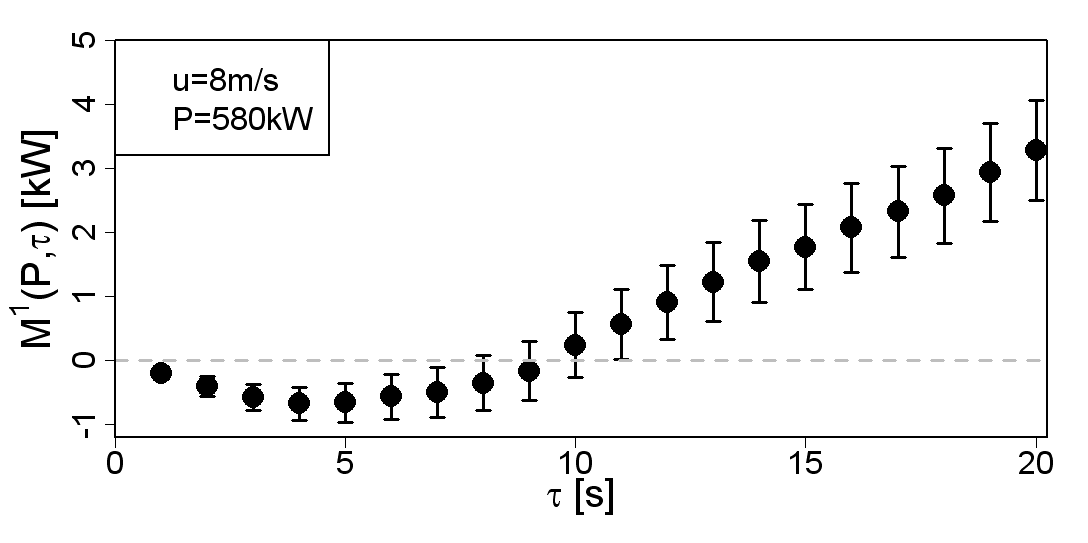}
      }
    \end{minipage}
    \begin{minipage}{0.95\textwidth}
      \centering{
        \hspace{0.01\textwidth}
        \includegraphics[width=0.45\textwidth]{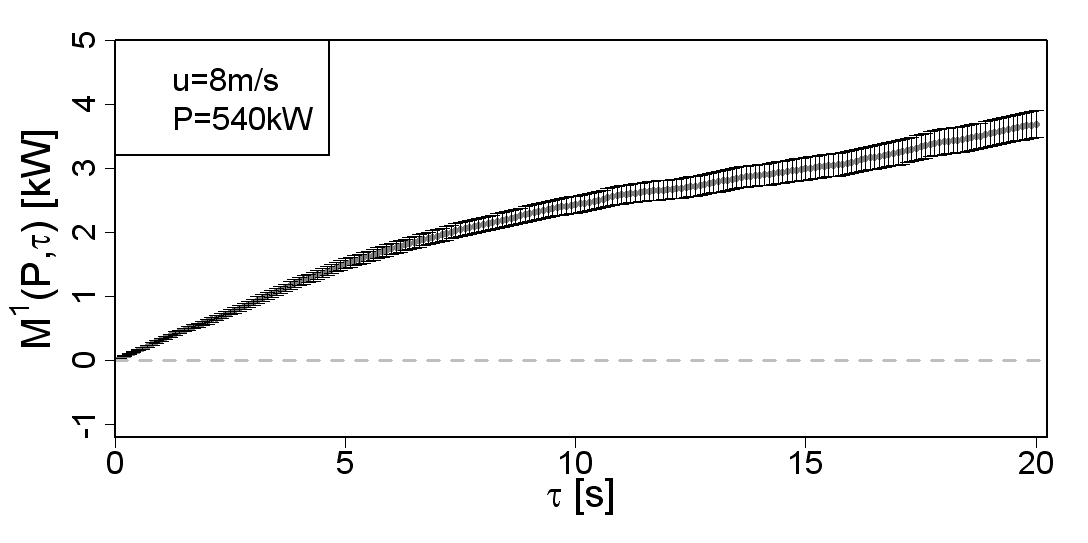}
        \hspace{0.05\textwidth}
        \includegraphics[width=0.45\textwidth]{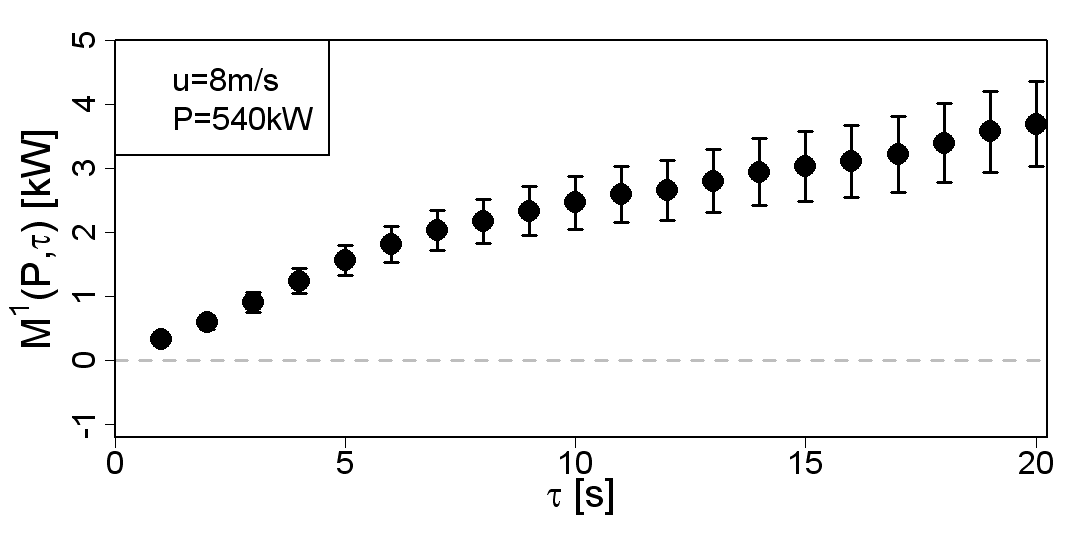}
      }
    \end{minipage}
  }
  \caption[Evolution of the conditional moments $M^{(1)}$ with increasing $\tau$ within two stacked power bins for different sampling rates]{\label{M1_1-200_u8} Evolution of the conditional moments $M^{(1)}$ with increasing $\tau$ 
    within two stacked power bins for different sampling rates of {\it data5}: (a), (c)~10\,Hz (original data); (b), (d)~1\,Hz (received from time averaging over 10 values).}
\end{figure*}
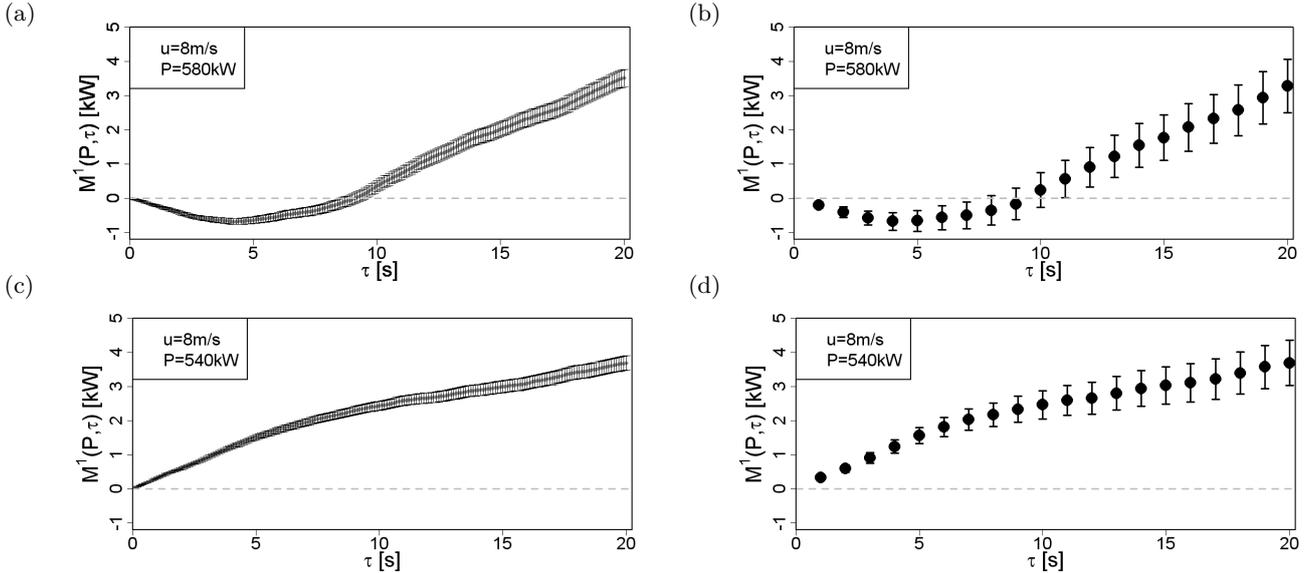

One central difference between ultrasonic and cup anemometer recordings is the temporal resolution.
For this reason, the impact of the sampling rate on the quality of the estimated LPC is of great relevance as this topic is related to the influence of the measurement device. 
For the representation of an ultrasonic measurement, we choose the 10\,Hz data from data set {\it data5} with a turbulence intensity of $I$=0.05.
The comparable data representing a cup anemometer are obtained by averaging the 10\,Hz data over 10 values. 
In this way, we obtain a time-averaged data set with a sampling frequency of 1\,Hz. 
These data are taken from the wind fields at the location of the hub height, as it is the case for a metmast measurement. 
Figure~\ref{M1_1-200_u8} shows the evolution of $M^{(1)}$ according to equation \eqref{eq:M1} with increasing $\tau$ for the two different sampling rates: 10\,Hz and 1\,Hz.
In principle, we have to quantify the quality of the sampling rate for all bins.
We show the results for two representative power bins (540$\pm$20\,kW and 580$\pm$20\,kW) within the wind speed bin of 8$\pm$0.25\,m/s. 
On the left side, the results for the ultrasonic related 10\,Hz data are illustrated. 
The bottom chart (c) shows the $M^{(1)}$ values for the power bin with the mean of 540\,kW; above chart (a) shows the results for the next higher power bin of 580\,kW.
Considering the first three seconds of $M^{(1)}$, a clear change in the sign of the slope of $M^{(1)}$ is seen by switching from one power bin to the other. 
Note that the drift term $D^{(1)}$ is given by the slope of $M^{(1)}(\tau)$, cf. equation~\eqref{eq:D1}. 
In the lower and upper power bin, the slope of a linear fit to $M^{(1)}$ within the first three seconds would be positive and negative, respectively.
Thus, this behaviour corresponds to a zero crossing of the drift function $D^{(1)}$ with a negative slope with respect to the power output, which describes a stable fixed point for equation~\eqref{langevin}.
In Figure~\ref{M1_1-200_u8}(b)+(d), the $M^{(1)}$ values for the same power bins are shown for the cup-related 1\,Hz data. 
It is clearly seen that the $M^{(1)}$ values for 1\,Hz data show the same evolution as the corresponding results for the 10\,Hz data in (a) and (c).

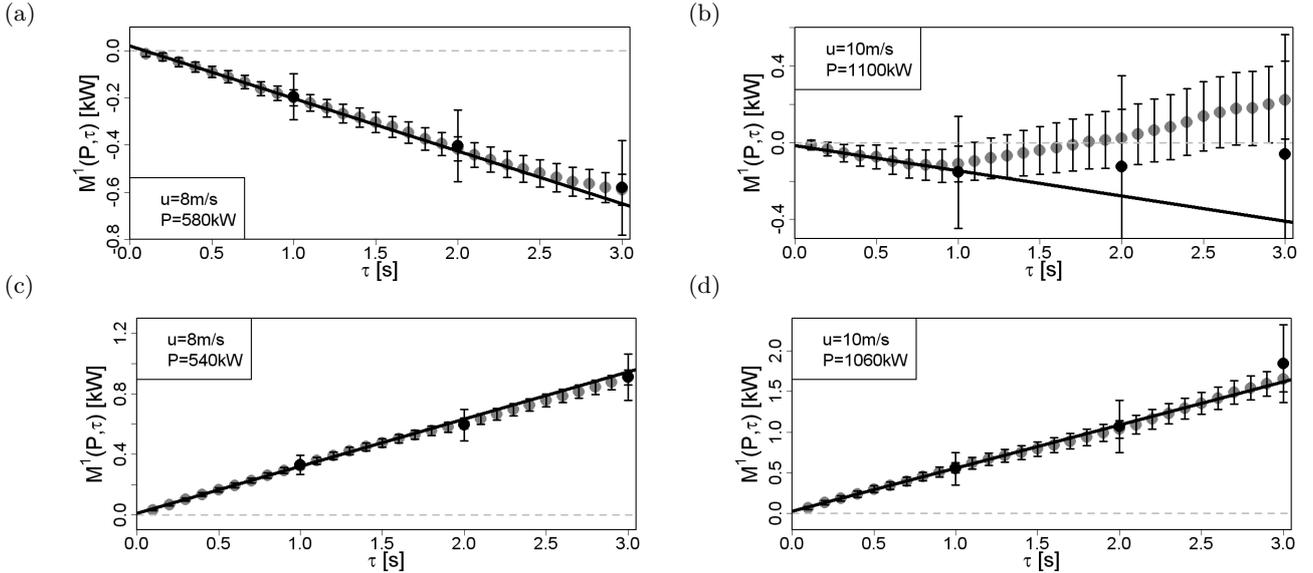
\begin{figure*}[htb]
  \unitlength 1.2mm
    \begin{picture}(0,0)
    \put(-4,15){(a)}
    \put(71,15){(b)}
     \put(-4,-15){(c)}
    \put(71,-15){(d)}
	\end{picture}%
\centering{
	\begin{minipage}{0.95\textwidth}
    \centering{	
		\includegraphics[width=0.45\textwidth]{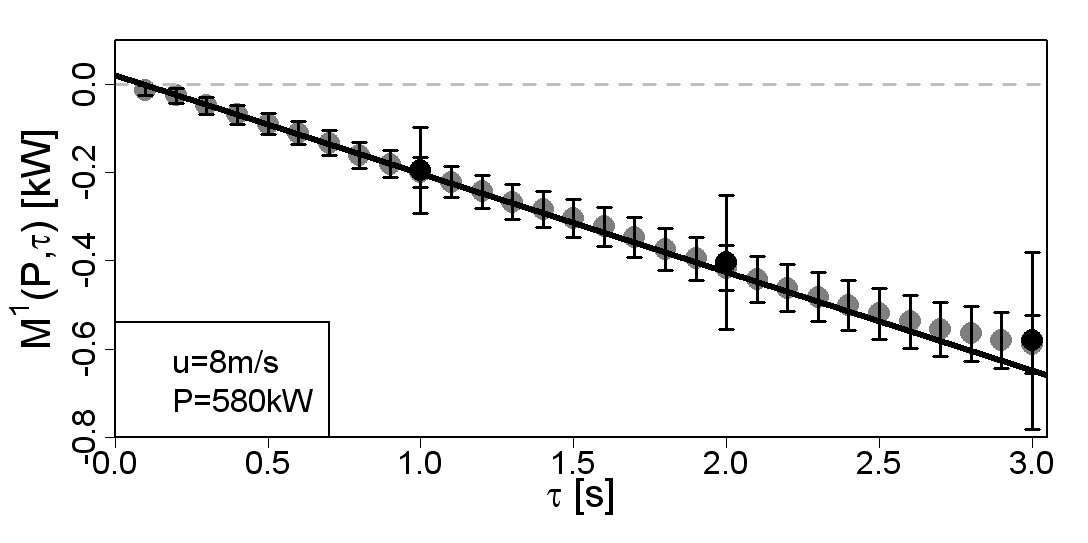}
		\hspace{0.05\textwidth}
		\includegraphics[width=0.45\textwidth]{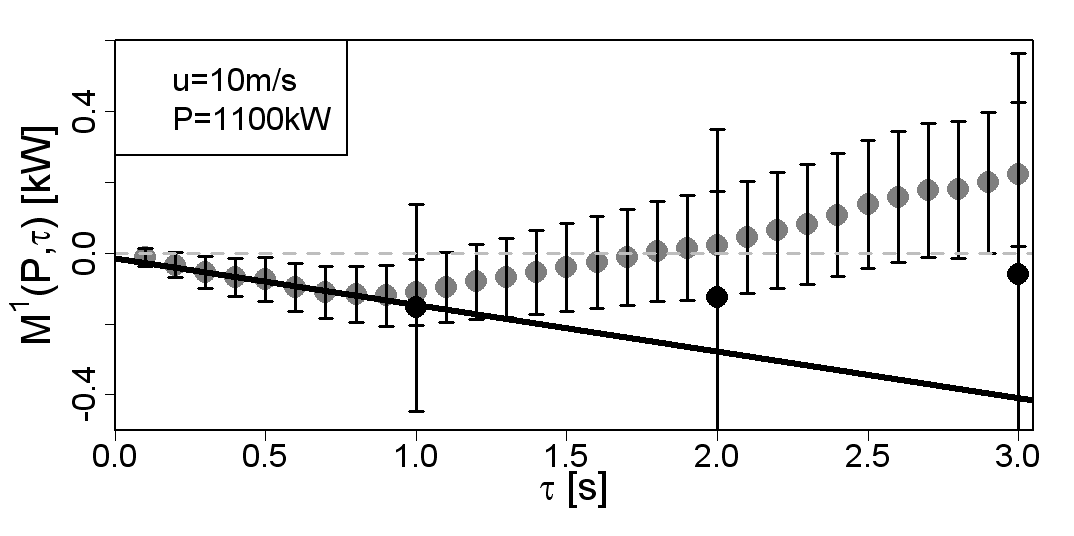}
		}
  \end{minipage}
  \begin{minipage}{0.95\textwidth}
    \centering{	
    \hspace{0.01\textwidth}
 		\includegraphics[width=0.45\textwidth]{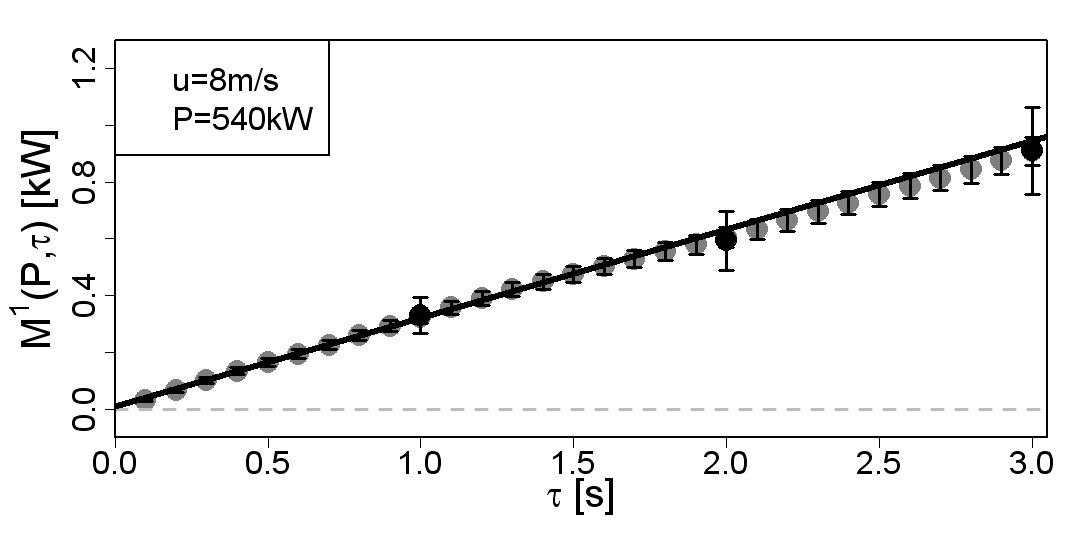}\hspace{0.05\textwidth}
		\includegraphics[width=0.45\textwidth]{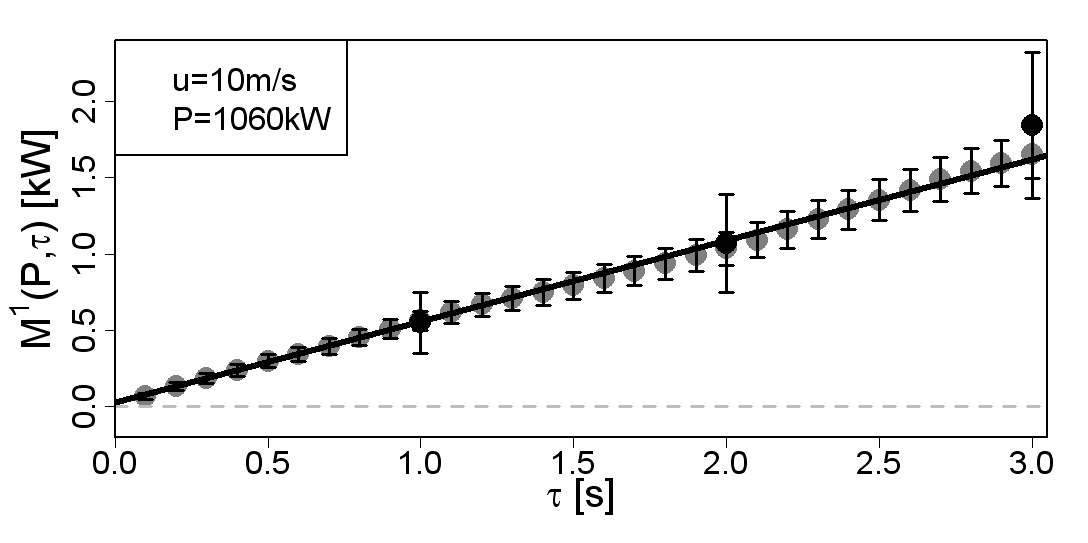}
		}
  \end{minipage}
 }
  \caption[Evolution of the conditional moments of {\it data5} within small $\tau$ values]{\label{M1_fit3-8} 
  Evolution of the conditional moments of {\it data5} within small $\tau$ values  
 for (a), (c)~u=8$\pm$0.25\,m/s (superposed zoom of Figure~\ref{M1_1-200_u8}) and (b), (d)~10$\pm$0.25\,m/s. The black line depicts the linear fit of the 10\,Hz data in the range of $\tau_{10} \in [0.3s,0.8s]$, respectively. }
\end{figure*}

In order to gain more detailed information about the slope of $M^{(1)}$, we take a closer look at the first seconds of the $M^{(1)}$ data, where the change in the algebraic sign takes place. 
In Figure~\ref{M1_fit3-8}(a) and (c) the first three seconds of the $M^{(1)}$ values from the diagrams of Figure~\ref{M1_1-200_u8} are displayed together. 
The grey dots are the values for the 10\,Hz, and the black ones those for the 1\,Hz data. 
In this presentation, the clear linear range of the $\tau$ values can be identified as $\tau_{10'} \in [0.3,1.5]$ for the 10\,Hz data and $\tau_1 \in [1,2]$ for the 1\,Hz data.
%
Figure~\ref{M1_fit3-8}(b) and (d) shows a similar presentation for another wind speed bin, namely u=10$\pm$0.25\,m/s. 
For this wind speed, which is close to the rated wind speed, the response dynamic is characterized due to very fast switches between the two different control strategies within partial load and full load operation of the wind turbine. 
This obvious fact is mirrored by the very short linear range for the $M^{(1)}$.
In order to represent as well the dynamics of this complex behaviour, the $\tau$ range has to be shortened to  $\tau_{10} \in [0.3,0.8]$.
In all four graphics, the black line depicts the linear fit of the $M^{(1)}$ values in this range for the 10\,Hz data. 
As the linear range of $M^{(1)}$ is longer for the 8\,m/s bin than for 10\,m/s bin, the calculation of the $M^{(1)}$ 
slope for u=8$\pm$0.25\,m/s is not affected because of the shorter $\tau_{10}$ range. 
The first two black dots of the 1\,Hz data are lying as well on the linear fit of the 10\,Hz data in Figure~\ref{M1_fit3-8}(a), (c) and (d). 
Thus, we obtain their similar results for the slope of $M^{(1)}$ with $\tau_1 \in [1,2]$.
The time-averaging procedure does not change the qualitative characteristics of the conditional moments.
A significant influence is only given in the certainty of the conditional moments as the smaller data base ($N$) leads to bigger error bars [see equation~\eqref{errorM}].
In the region of transition to full load, it is difficult to catch the complex dynamics with both 10\,Hz and 1\,Hz data [Figure~\ref{M1_fit3-8}(b)]. 
Especially for the 1\,Hz data, the uncertainty increases in this region, and longer simulation periods are needed. 

\begin{figure*} 
  \unitlength 1.2mm
    \begin{picture}(0,0)
    \put(-4,15){(a)}
    \put(71,15){(b)}
     \put(-4,-25){(c)}
    \put(71,-25){(d)}
	\end{picture}%
\centering{
	\begin{minipage}{0.95\textwidth}
    \centering{	
		\includegraphics[width=0.40\textwidth]{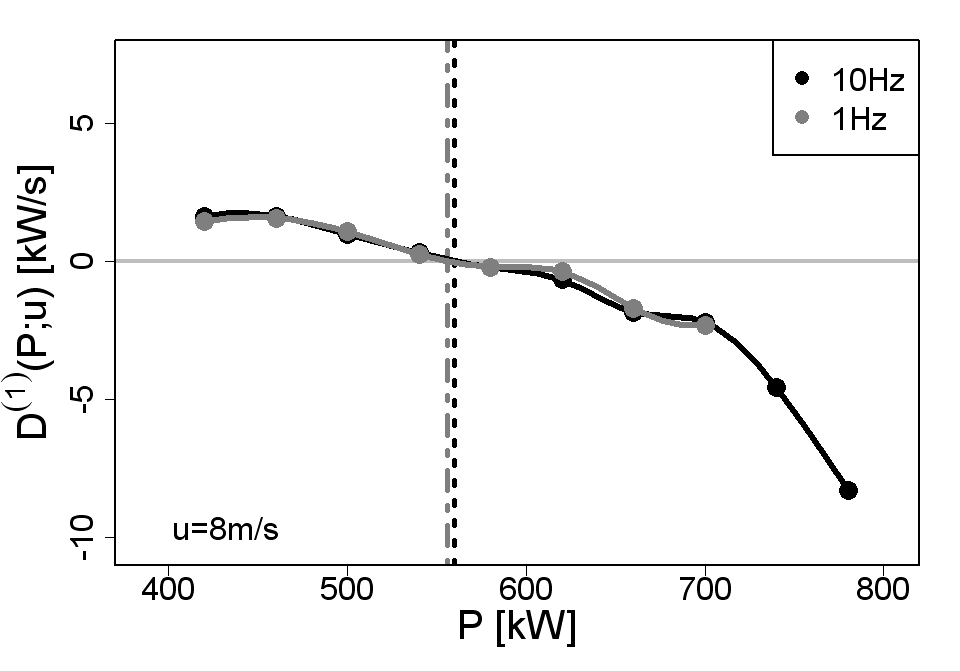}
		\hspace{0.1\textwidth}
		\includegraphics[width=0.40\textwidth]{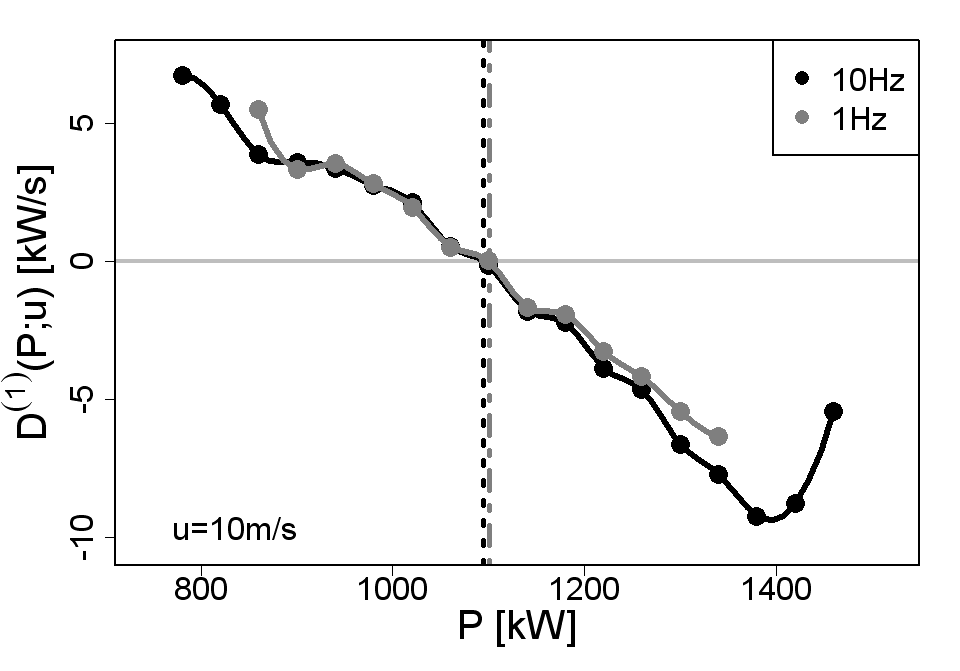}
		}
  \end{minipage}
  \begin{minipage}{0.95\textwidth}
    \centering{	
		\hspace{0.005\textwidth}
 		\includegraphics[width=0.40\textwidth]{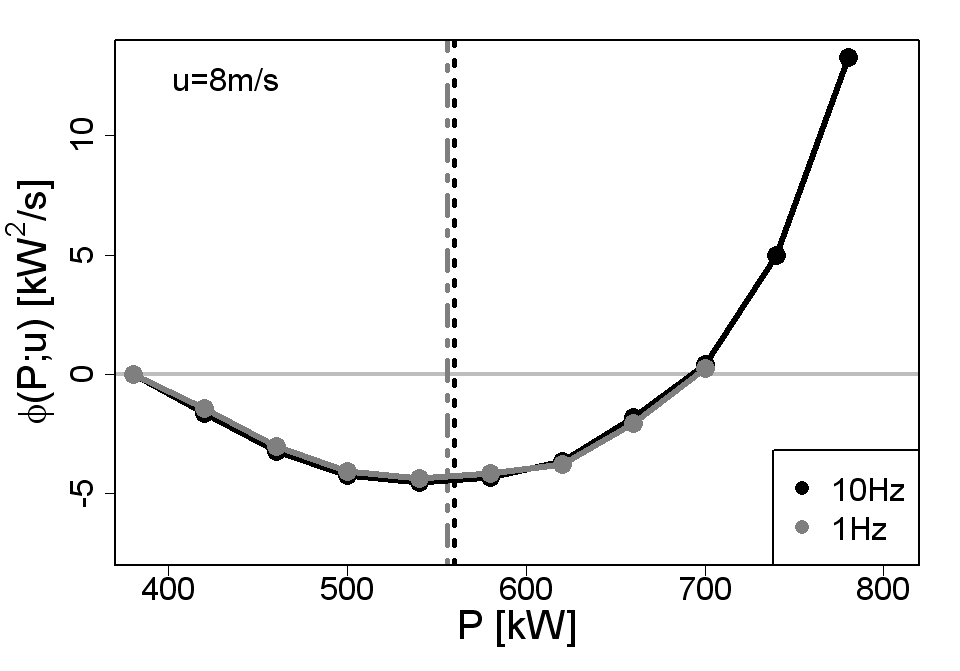}
	\hspace{0.1\textwidth}
		\includegraphics[width=0.40\textwidth]{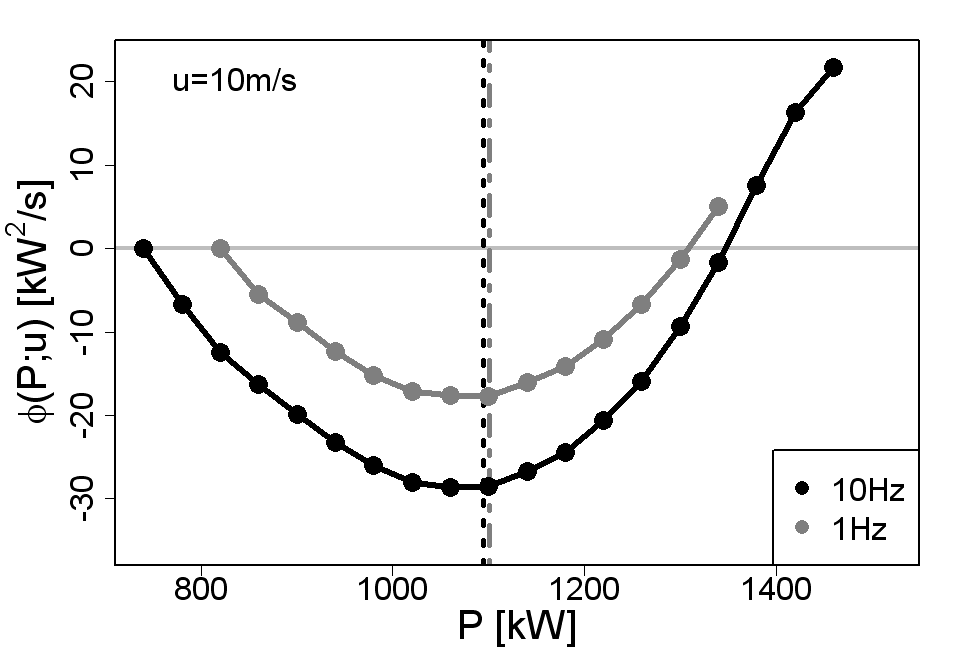}
		}
  \end{minipage}
 }
  \caption[Drift function $D^{(1)}_i(P;u)$ for $u$=8$\pm$0.25\,m/s and $u$=10$\pm$0.25\,m/s]{\label{D1_10-1} Drift function $D^{(1)}_i(P;u)$ for (a)~$u$=8$\pm$0.25\,m/s and (b)~$u$=10$\pm$0.25\,m/s. The corresponding potentials $\Phi(P;u)$ are given in (c) and (d). In all figures, the 10\,Hz data are given in black, and the 1\,Hz data in grey. The dots display the estimated results per power bin. The vertical lines depict, in all four figures, the zero crossing of the drift function, which is the estimated fixed point. The black dotted line gives the fixed point for the 10\,Hz data, and the grey dash dotted line for the 1\,Hz data. }
\end{figure*}
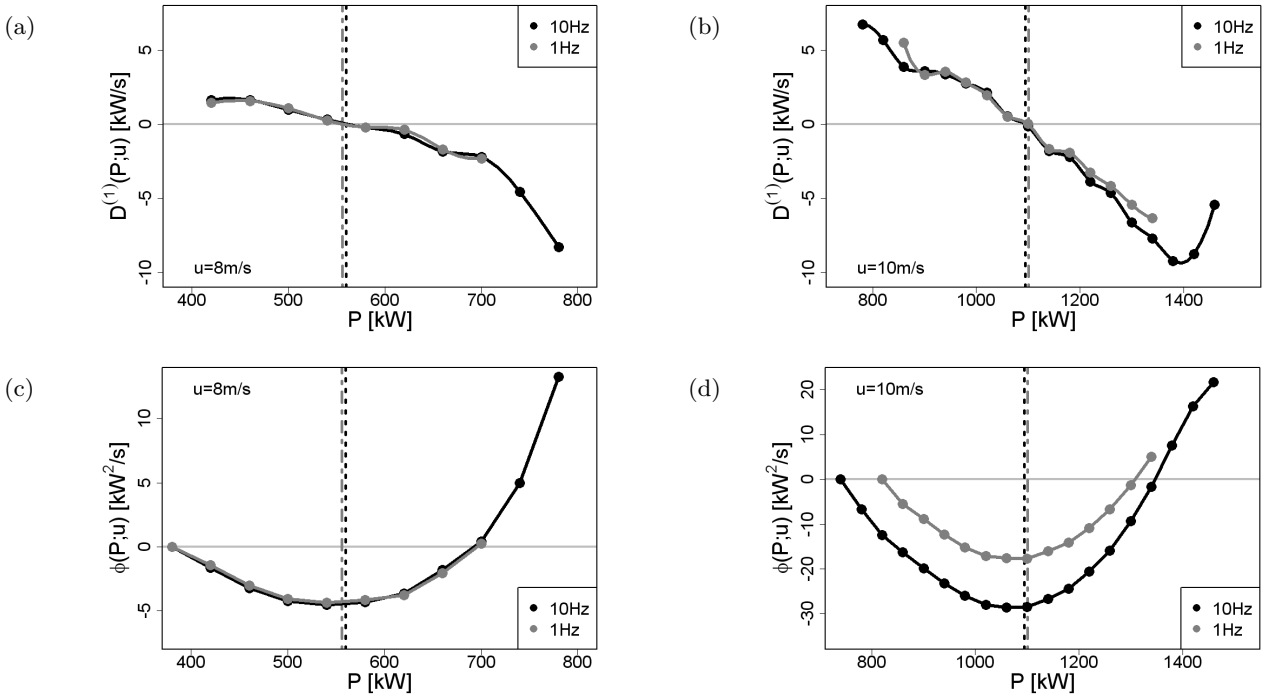

The estimated drift coefficients $D^{(1)}_i(P;u)$ are shown in Figure~\ref{D1_10-1}(a) and (b) for the wind speed bins 8 and 10\,m/s as a function of different $P$-values. 
Each point displays the drift coefficient estimated from the slope of the linear fit of the conditional moments $M_{ij}^{(1)}$.
For the fit of all estimated drift coefficients as a function of the power output, a cubic spline is used.
This procedure is in correspondence with the findings of Gottschall~{\it et al.}~\cite{Gottschall2008b}.  
They showed that symmetries in $D^{(1)}$ are of extreme importance for the robustness of fixed points especially in the presence of measurement noise and for finitely resolved processes. 
Since the fixed point analysis is a local linear method, it is affected if and only if the disturbance brings the  dynamics out of the linear vicinity of the fixed point.
Thus, it is advisable to interpolate the $D^{(1)}$ with piecewise polynomials to identify  the zero crossing in the nearly linear range.
Obviously, the calculated zero crossings $P_L$ with a negative slope of $D^{(1)}$ are nearly identical for the 1\,Hz and the 10\,Hz cases [see Figure~\ref{D1_10-1}(a) and (b)]. 
The estimated power values are 560.0$\pm$5.1\,kW (10\,Hz) and 555.8$\pm$6.0\,kW (1\,Hz) within the wind speed bin of $u$=8$\pm$0.25\,m/s and 1094.3$\pm$13.1\,kW (10\,Hz) and 1100.6$\pm$14.6\,kW (1\,Hz) within the wind speed bin of $u$=10$\pm$0.25\,m/s.
It is also seen that the drift function $D^{(1)}_i$ has a clear linear range.
This behaviour is expected as the deterministic part of the Langevin equation can be written as
\begin{equation}\label{eq-d1}
D^{(1)}(P;u)=- \alpha [P(t)-P_L(u)] \quad,
\end{equation}
where $P_L$ denotes the fixed point of the power output $P(t)$.
This term describes the simplest relaxation case of the instantaneous $P(t)$ with an exponential decay or grow on sudden changes.
The noise-free solution for equation~\eqref{langevin} with such a $D^{(1)}$ is then 
$[P(t)-P_L(u)] \propto e^{\alpha \cdot  t}$.
In the simplest case, the relaxation factor $\alpha$ is  a constant and may also be a function of wind speed and power output~\cite{rauh2004a, rauh2004b}.


A rather intuitive presentation of the calculated fixed points is given with the potential 
\begin{equation}\label{pot}
\Phi(P;u)=-\int^P D^{(1)}(P';u)dP'   \quad .
\end{equation}
In this presentation, a stable fixed point can be identified as a minimum value of the potential $\Phi$(P;u).
In Figure~\ref{D1_10-1}(c) and (d), the potentials of the drift functions are shown.
The different absolute values of the potentials in Figure~\ref{D1_10-1}(d) contain no significant information.
Because of the potential's definition [equation~\eqref{pot}], the valley is shifted in y-direction if more estimated drift coefficients are available for low power bins.
The vertical lines depict the stable fixed points (zero crossing of the drift function): black dotted for the 10\,Hz data and grey dash dotted for the 1\,Hz data.
The location of a fixed point calculated from $D^{(1)}_i$ is  always a little bit shifted to the right (i.e. to higher $P$ values) from the visual minimum of the potential. 
This is due to the definition of the potential and the fact that the zero crossing of the drift function is mostly between two calculated coefficients. 
With a higher resolution of power bins, the right shift in the potential vanishes.
In real life, a higher resolution of power bins is not desirable as this is directly connected with the need of a higher amount of data.
Thus, we emphasize to use the drift function for the calculation of the fixed points and the minimum of the potential only for visualization. 
This advise is supported by the findings presented in previous work~\cite{Gottschall2008b}.

\begin{figure*} 
  \unitlength 0.95mm
  \centering
  \begin{picture}(0,0)
    \put(-4,28){(a)}
    \put(71,28){(b)}
    \put(-4,-35){(c)}
    \put(71,-35){(d)}
  \end{picture}%
        \includegraphics[width=0.3\textwidth]{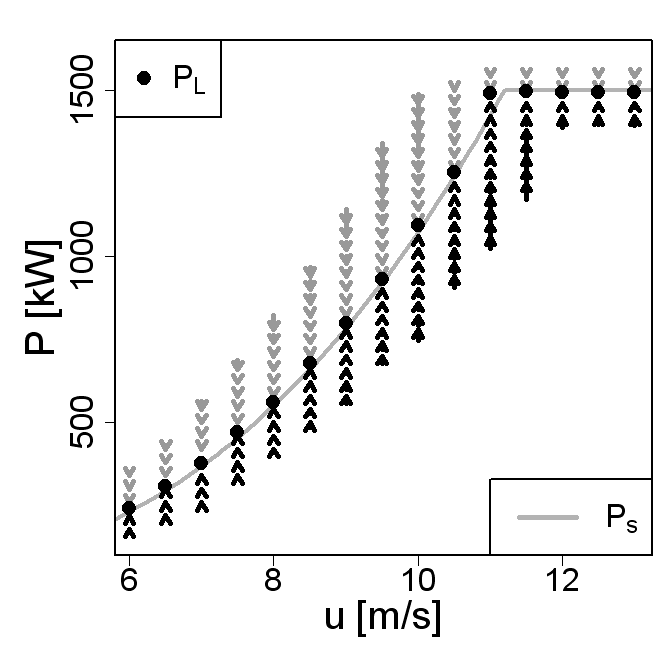}
        \hspace{0.1\textwidth}
        \includegraphics[width=0.3\textwidth]{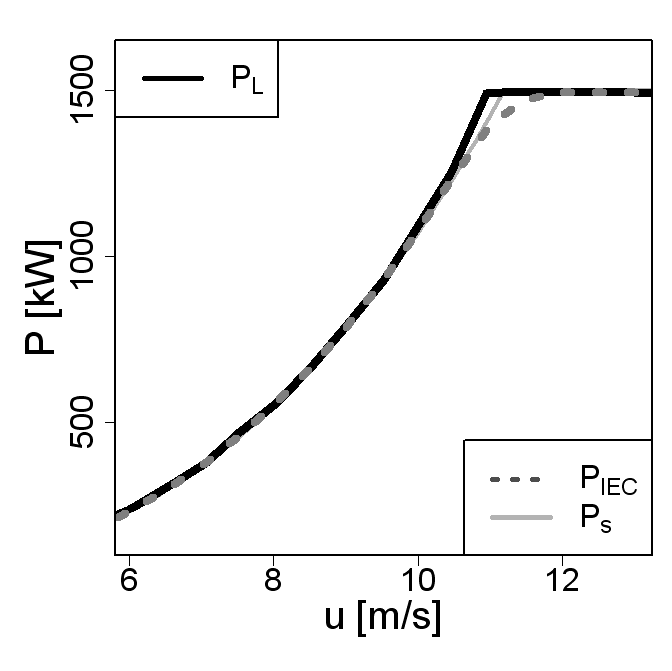}

        \includegraphics[width=0.3\textwidth]{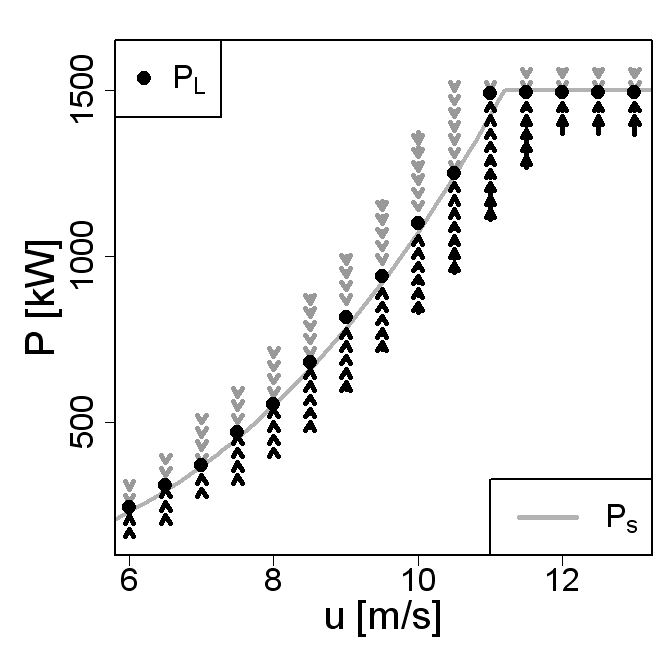}
        \hspace{0.1\textwidth}
        \includegraphics[width=0.3\textwidth]{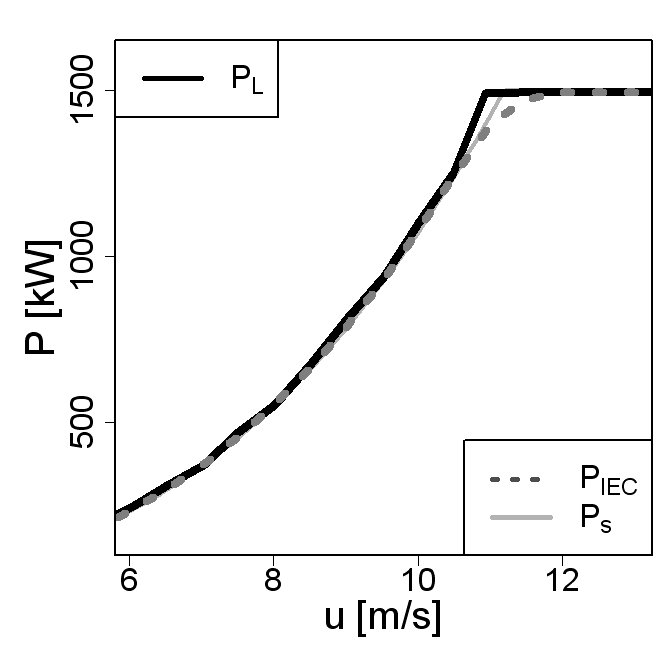}
  \caption[Drift field and LPC for the data sets of {\it data5} with 10\,Hz and 1\,Hz sampling rates.]{\label{LPC_beide}  Drift field (left) and LPC (right) for the data sets of {\it data5} with (a), (b)~10\,Hz and (c), (d)~1\,Hz sampling rates. The slope estimation for each bin is carried out with $\tau_{10} \in [0.3,0.8]$ and  $\tau_{1} \in [1,2]$, respectively. 
  }
\end{figure*}
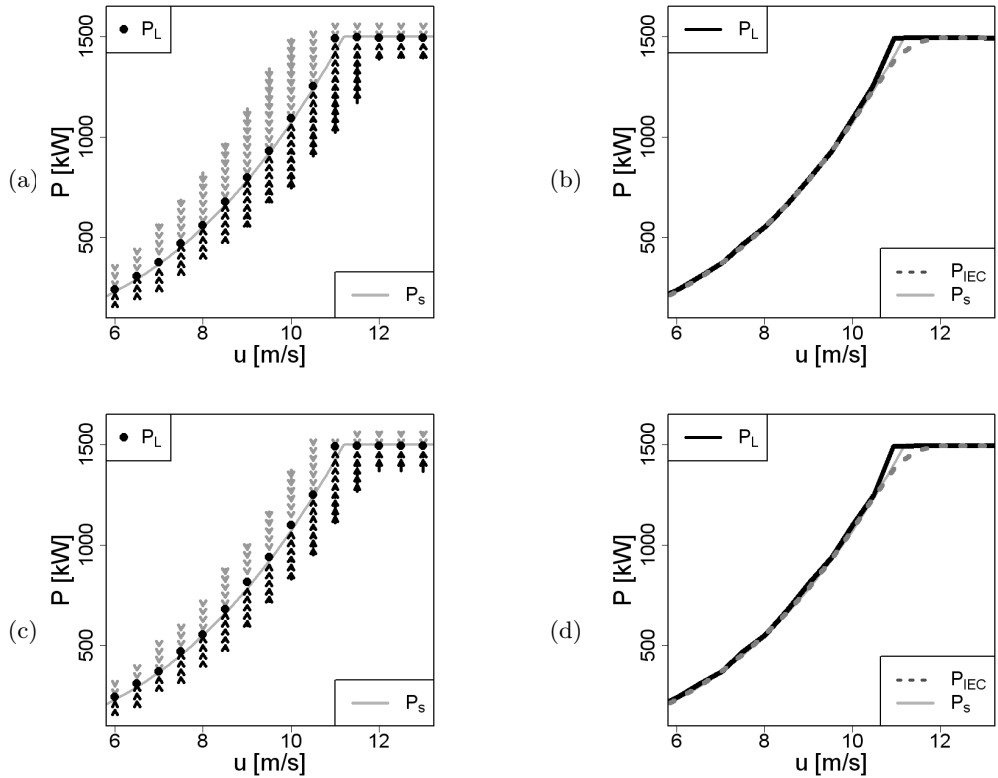

In order to finish our investigation about the different anemometer types, we calculate the LPC for the two different sampling rates.
In Figure~\ref{LPC_beide} the drift field (a), (c) and LPC (b), (d) of {\it data5} are displayed for the two sampling rates, respectively.
In (a) and (c), the black dots are displaying the fixed points $P_L$ within one wind speed bin, where $D^{(1)}$=0 and the slope of the drift coefficients is negative. 
As the analysis is based on a numerical model, we are able to determine the power values $P_S$ from simulations with a constant uniform (homogeneous and laminar) inflow.
After a short time this noise free system relaxes to a steady-state $P_S$ according to equation~\eqref{eq-d1}. 
These steady-states $P_S$ are depicted by the solid grey line. 
A visualisation of equation~\eqref{langevin} can be achieved by plotting the values of matrix $D^{(1)}$  using arrows.
The size of the arrows is scaled by the values of $D^{(1)}$, and the colors indicate the direction of the dynamics.
A black arrow represents a positive value of $D^{(1)}$, indicating that, in this region, the power output tends to increase. 
A grey arrow illustrates the opposite case. 
The power output is driven by the controller of the WEC to a lower power level, and $D^{(1)}$ is negative. \\
%
For both data sets, we could successfully calculate the corresponding LPC $P_L(u)$.
The mean error of the estimated fixed points is below 1\% (0.7\% for the 10\,Hz data and 0.9\% for the 1\,Hz data).
Thus, the difference between both results is negligible.
The drift field of the 1\,Hz data is slightly more compact in comparison with the one of the 10\,Hz data. 
This is caused by the averaging of the data, as the fluctuations are damped and the dynamics is truncated to a smaller range. Nevertheless, the characteristic dynamics of the system coincides for both sampling rates.
Furthermore, the mean root square deviation of the estimated fixed points for the 10\,Hz in comparison with the time averaged 1\,Hz data is only 0.6\%.
As expected, most of the estimated fixed points correspond with the steady states of the wind turbine [see Figures~\ref{LPC_beide}(b) and (d)].
Driven by the turbulence-induced stochastic noise, the system should relax to the noise-free solution.
This is the case for the operation at partial and full power.
In the transition range, the power conversion system is disturbed. 
In this range, the fixed points may deviate from the steady states.\\
An IEC power curve based on the data is estimated as well, following the proposed procedure~\cite{IEC-12} with a wind speed bin width of 0.5\,m/s.
In order to compare both methods over the whole range of wind speeds, the LPC is presented as a black curve in Figure~\ref{LPC_beide}(b) and (d) and is displayed together with the IEC curve (dashed curve). 
In the transition range to full load, the difference between the IEC and the LPCs become obvious.
The LPC describes the dynamical response of the wind turbine and is able to display the initiation of different control strategies as well. 
Thus, the fixed point for 11$\pm$0.25\,m/s is already in the range of full load, as the turbine is working here mostly under full load condition due to the turbulent inflow. 
The IEC curve underestimates the power performance in this region because of systematic errors caused by the temporal averaging~\cite{Gottschall2008,Milan2009}. 

Summing up, it seems to make no significant difference which anemometer type, or more precisely, which temporal 
resolution is used to obtain a reliable LPC. 
Most of the power bins show a similar characteristics, and the linear fit of $M^{(1)}$ has the same slope.
Furthermore, on average, the estimated fixed points are identical within a range of 0.6\%. 
The Langevin approach takes the short-time dynamics into account and can display the relation of wind and power in a direct way. 
The method is not restricted by the limitations of a linear averaging method as the IEC power curve.
Nevertheless, if  details of the control system are of interest, a 1\,Hz sampling rate is not advisable.
The crucial dynamics happen within some seconds, and a slope estimation based on two points is in general not reliable (just the algebraic sign of the slope can be certainly obtained).

\subsection{Spatially averaged data}\label{lidar}
Nowadays, measurement devices based on remote sensing technology are very popular.
A big advantage of those devices is the mobility and the possibility to measure also the wind speeds at higher altitudes in comparison with a fixed installed metmast.  
In the, e.g. Leosphere Windcube LIDAR, an infrared laser beam is inclined by approximately 30\degree against the vertical direction and takes beamwise Doppler measurements of the wind velocity. 
The operation principle of the Leosphere Windcube is, for instance, described in reference~\cite{waechter_dewek, waechter2008}.
In contrast to measurements with cup and ultrasonic anemometers, the wind velocity is not measured at one location but averaged over a horizontal range. 
Additionally, the three-dimensional wind vector is only derived from the most recent measurements during a certain time period, depending on the time of circulation of the laser beam.
The Leosphere WindCube LIDAR performs a distributed measurement over a large spatial area and does not reach the temporal resolution of the current cup and ultrasonic anemometers. 
Because of its measurement principle, the Windcube LIDAR achieves only a sampling rate of 0.67\,Hz.

In this section, we want to investigate if the Langevin approach is in general working with spatially averaged wind measurements as it is characteristic for LIDAR data.
It is not our aim to model the LIDAR measurement in every detail but to give an impression of the extent to which the spatial averaging of measured data influences the Langevin approach. 
In order to represent a LIDAR measurement over a large spatial area, we create a temporal and horizontal averaged data set from high frequent synthetic data.
The already investigated synthetic data of {\it data5} with a sampling rate of 10\,Hz are used as data basis for the averaging procedure.
As is customary, we take the averaged wind speed at hub height (84\,m) as underlying wind speed for the power curve calculation.  
The wind field grid consists of nine points at 84\,m hub height. 
The following simple averaging procedure is applied, which gives a rough approximation to how a LIDAR works:
at first, the nine time series of the grid points at hub height are averaged over every second to 1\,Hz data.
Then, the horizontal mean value is calculated for every second from the nine timely averaged values at hub height. 
This procedure leads to a LIDAR-like data set with a spatially averaged inflow. 
In order to match the sampling rate, the calculated power output is also averaged over every second from 10\,Hz to 1\,Hz data. 

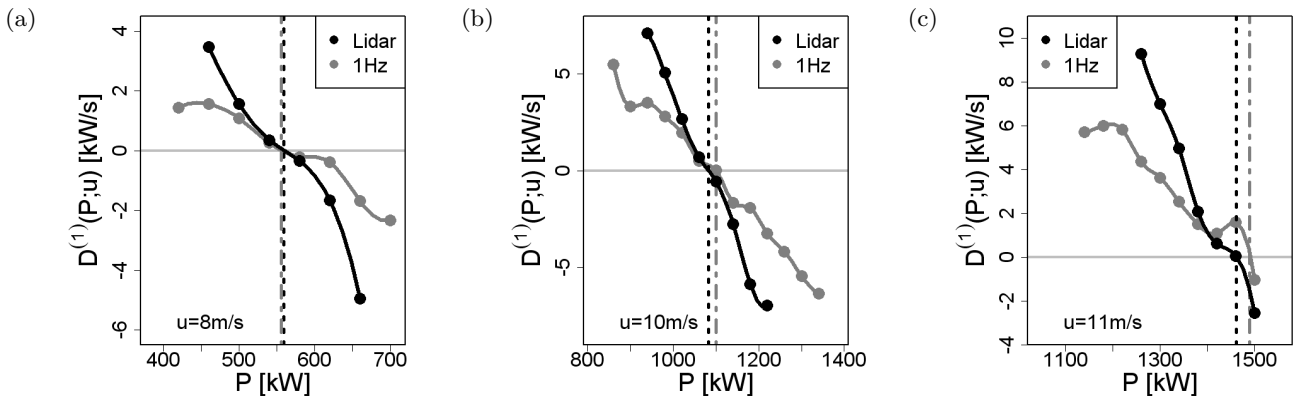
\begin{figure*} 
  \unitlength 1.2mm
  \begin{picture}(0,0)
    \put(-4,20){(a)}
    \put(46,20){(b)}
    \put(95,20){(c)}
  \end{picture}%
  \centering{
    \begin{minipage}{0.95\textwidth}
      \centering{
        \includegraphics[width=0.28\textwidth]{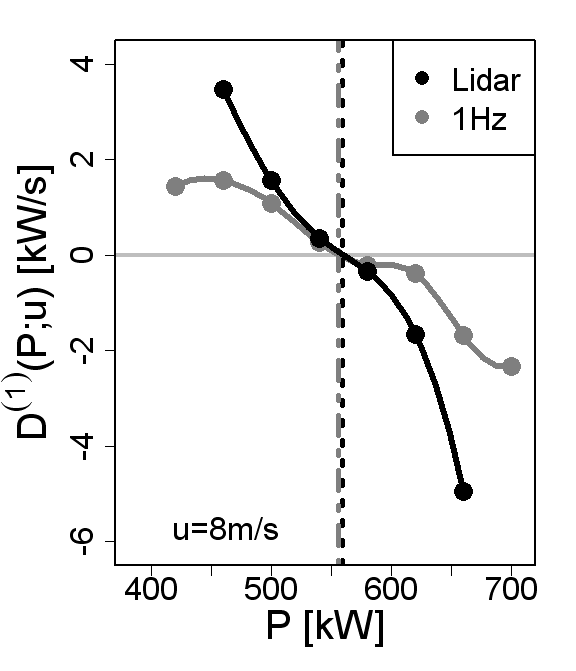}
        \hspace{0.05\textwidth}
        \includegraphics[width=0.28\textwidth]{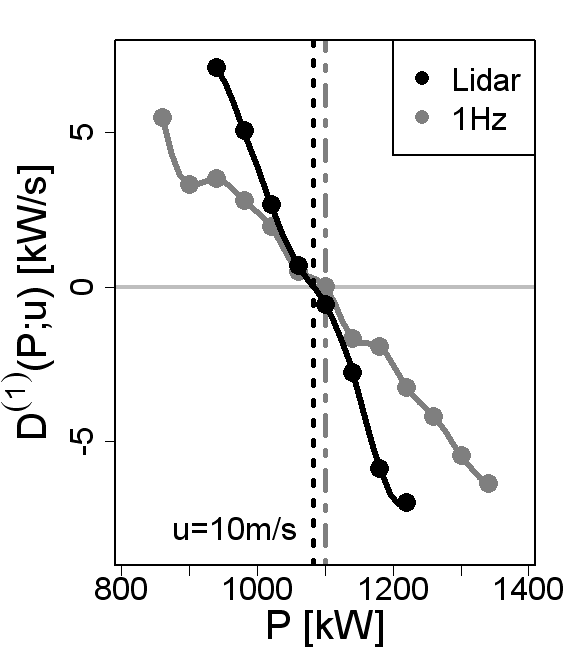}
        \hspace{0.05\textwidth}
        \includegraphics[width=0.28\textwidth]{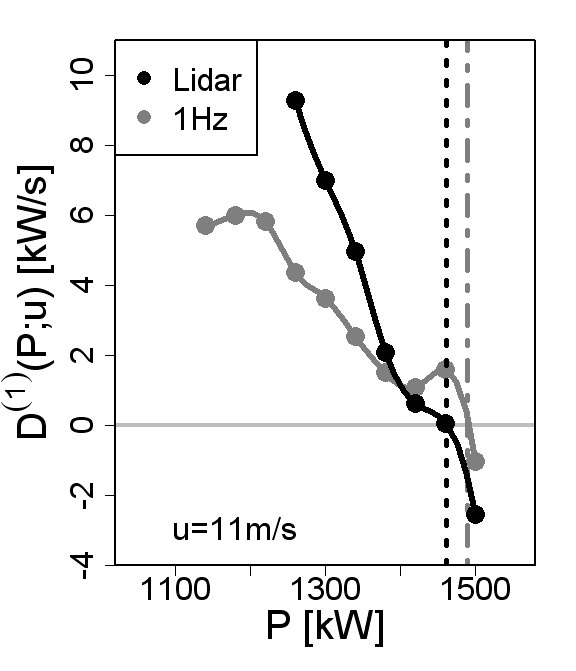}
      }
    \end{minipage}
  }
  \caption[Drift functions $D^{(1)}(P;u)$ for the the modelled LIDAR data set in comparison with the 1\,Hz data set at hub height]{\label{D1_all} Drift functions $D^{(1)}(P;u)$ for the the modelled LIDAR data set in comparison with the 1\,Hz data set at hub height. The {\it data5} were used within the wind speed bins (a)~8.0$\pm$0.25\,m/s, (b)~10.0$\pm$0.25\,m/s and (c)~11.0$\pm$0.25\,m/s. }
\end{figure*}

The sampling frequency of the 1\,Hz LIDAR-like data differs slightly from the 0.67\,Hz sampling rate of the Windcube LIDAR. 
Nevertheless, we take the sampling rate of 1\,Hz for the further analysis.
The data set {\it data5} with a sampling frequency of 1\,Hz [see section~\ref{freq}] has the same power output as the LIDAR-like data and differs only in the processing of the underlying wind speed.
By comparing both data sets, we are able to identify influences on the estimated LPC caused solely by the spatial averaging procedure.
The estimated drift functions for the LIDAR-like data are displayed in Figure~\ref{D1_all} for some selected wind speeds.
For comparison, the results of the 1\,Hz data of the corresponding wind speed measurements at hub height are displayed too. 
The 10\,Hz data showed similar results as the 1\,Hz data and were not included in the figure for clarity.
The comparisons show clear evidence that the drift functions of the spatially averaged LIDAR data set deviate significantly from the one-point measurements.
Nevertheless, the fit of the estimated drift coefficients lead almost in every case to slightly lower fixed points. 

\begin{figure*} 
  \unitlength 0.9mm
  \centering
  \begin{picture}(0,0)
    \put(-4,28){(a)}
    \put(71,28){(b)}
  \end{picture}%
        \includegraphics[width=0.3\textwidth]{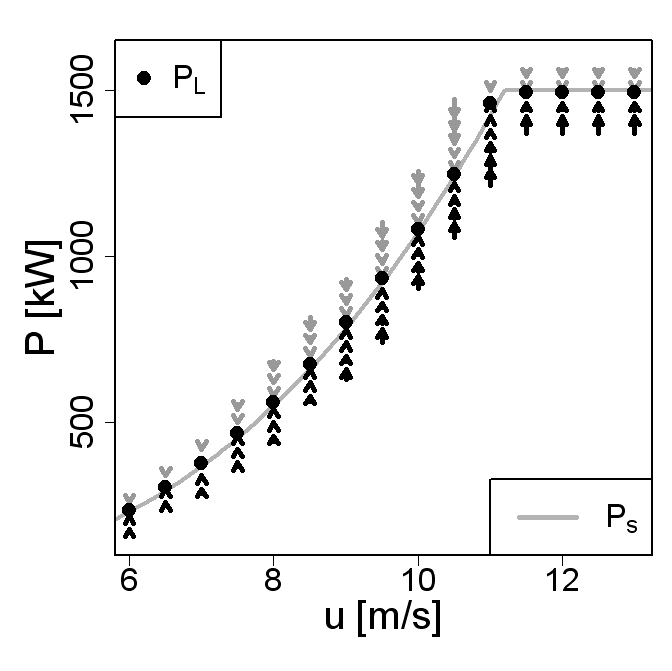}
        \hspace{0.1\textwidth}
        \includegraphics[width=0.3\textwidth]{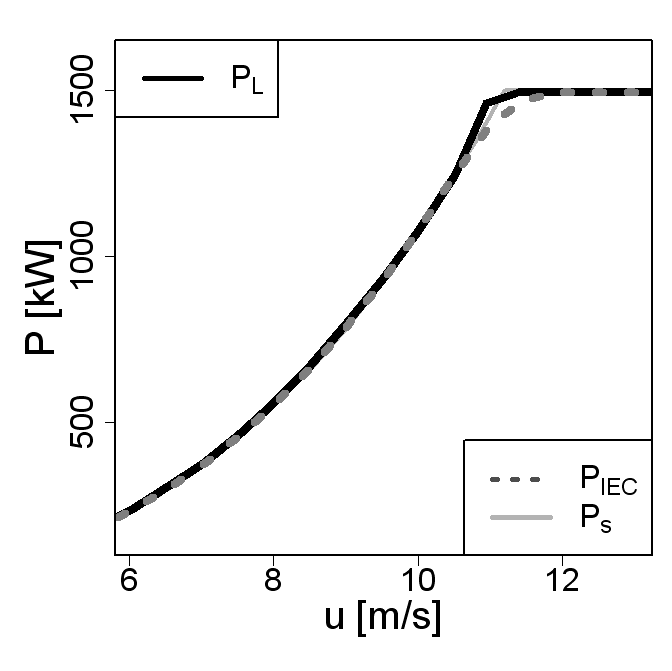}
  \caption[Drift field and Langevin power curve for the modelled LIDAR-related data]{\label{LPC_Lidar} (a)~Drift field and (b)~Langevin power curve $P_L$ for the modelled LIDAR-related data in comparison with the common IEC power curve $P_{IEC}$ (dashed line). In both graphs the solid grey line depicts the steady states $P_s$ of the WP\,1.5 wind turbine.
The fixed points of the conversion process are displayed in black as (a)~points and (b)~solid line. Data basis for the analysis is the {\it data5} data set after horizontal averaging. The slope estimation for each bin is carried out with $\tau_{1} \in [1,2]$. }
\end{figure*}
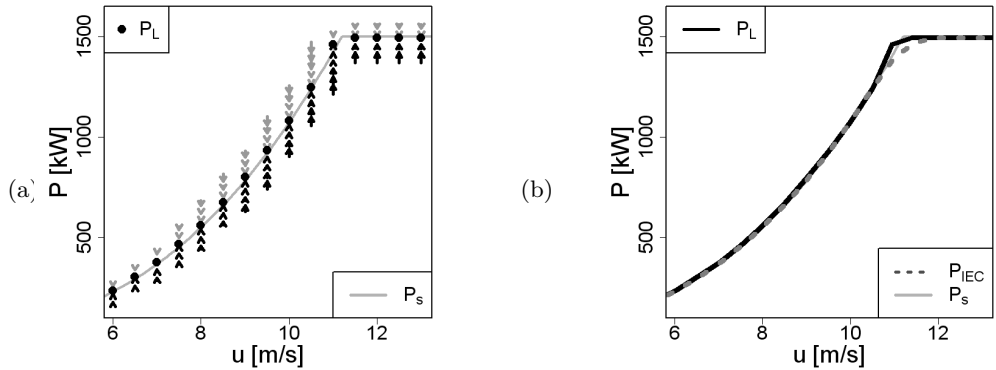

Obviously, the very fast dynamics of the control system cannot be reflected within the wind speed bin of 11$\pm$0.25\,m/s for the spatially averaged data.
The tendency of the control system to switch into a higher state is not displayed, and the fixed point (zero crossing of the drift function) is located around 50\,kW lower.
Let us now take a closer look on the complete LPC.
The drift field and calculated fixed points for the LIDAR-related data are shown in Figure~\ref{LPC_Lidar}.
The drift field is once again presented on the left-hand side~(a), and the LPC on the right-hand side~(b). 
The graphs are prepared according to the description of section~\ref{freq}.
The horizontal averaging procedure has a strong damping effect on the fluctuations as expected.
This fact is mirrored in a more compact drift field in comparison to the original data. 
Nevertheless, the drift field and the fixed points coincide very well with the results in Figure~\ref{LPC_beide}(a) and (b). 
With the exception of the fixed point for the wind speed bin of 11\,m/s, all fixed points are almost similar.
They coincide with the 1\,Hz data within a mean deviation of 1.1\%.
All fixed points of the wind turbine can be obtained with a high certainty as the mean error is 0.6\%.

In conclusion, we can state that the Langevin approach can be performed on the basis of LIDAR measurements and that the steady states of the wind turbine can be calculated with a high certainty.
But using a spatially averaged wind speed as input for the Langevin approach results in a reduced area of the drift field in the phase space. 
The dynamics on short time scales, like control strategies of the turbine, cannot be grasped due to the spatial averaging.
This restriction becomes more pronounced if the sampling rate of the LIDAR-like data is additionally less than 1\,Hz.
Possibilities to overcome this drawback are either in operating the LIDAR in an other mode or to combine results from the LIDAR measurements with according results from a hub anemometer. 
Based on the hub anemometer, a quite accurate relative LPC can be estimated, but errors are given due to the systematic errors in the absolute values of the wind speed. 
These errors caused by the blockage of the rotor are not present in the LIDAR data.

\section{Verification of the applicability of the Langevin power curve}
In the last section, we have investigated how reliable the LPC can be determined from differently measured wind data. 
Up to now, the same amount of data has been used. 
Next, we will discuss the important issue of how many data sets are needed in order to get reliable results.
Based on these insights, we proceed to study the influence of the site-specific turbulence intensity on the LPC.
It has been claimed several times that the LPC should be site independent.  
Thus, we investigate the impact of the changing turbulence degree on the LPC by using the numerical model of the WP\,1.5 wind turbine.
The last aspect of this section is devoted to the ability of the method to monitor the condition of the power conversion process.

 
\subsection{Local criteria for reliable slope estimates}\label{slope}
The calculated LPCs in the former section are based on a data set of 105\,h.
We have seen that  we get reliable results for the fixed points.
Next, we want to quantify the amount of data, which is at least necessary to get reliable estimates.
In the case of field measurements, this issue is related to the length of a measurement campaign.

\begin{figure*} 
  \unitlength 1mm
    \begin{picture}(0,0)
    \put(-4,15){(a)}
    \put(85,15){(b)}
	\end{picture}%
        \centering{
          \begin{minipage}{0.95\textwidth}
            \centering{	
              \includegraphics[width=0.45\textwidth]{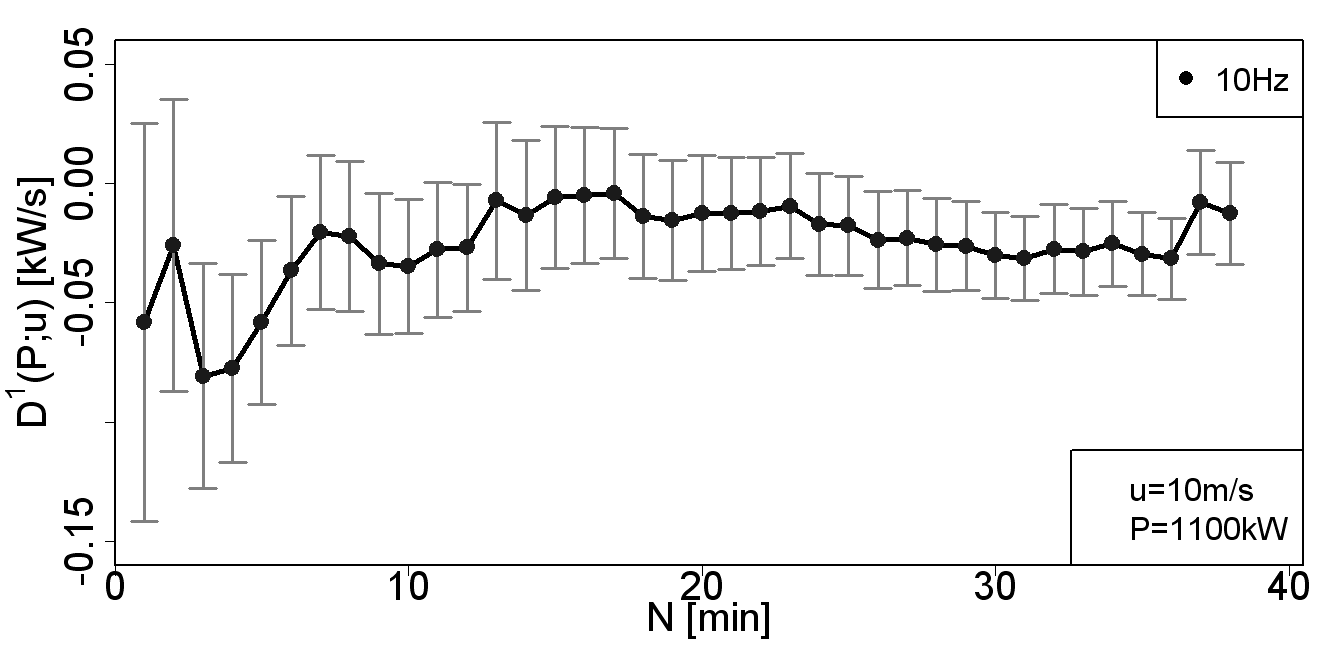}
              \hspace{0.05\textwidth}
              \includegraphics[width=0.45\textwidth]{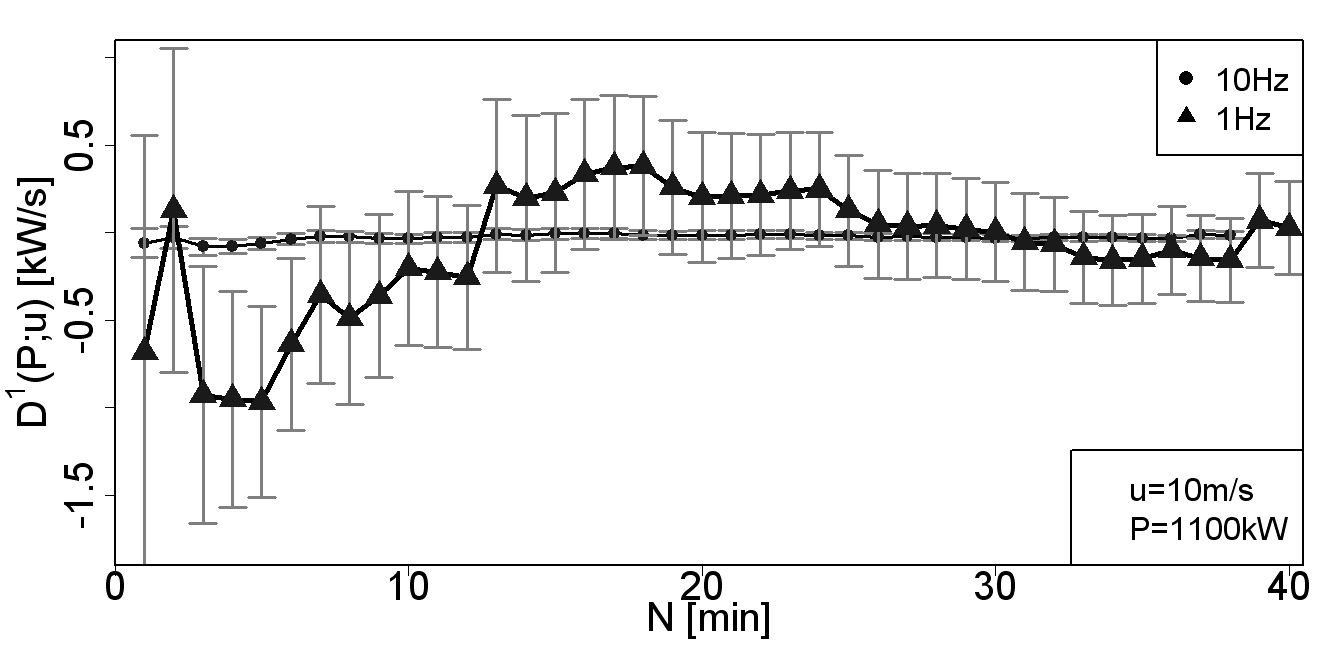}
            }
          \end{minipage}
        }
        \caption[Evolution of the drift coefficient in dependence of the length (in measuring time) of the data basis]{\label{D1_vgl}
          Evolution of the drift coefficient in dependence of the length (in measuring time) of the data basis. (a)~estimates for the 10\,Hz data of {\it data5}, (b)~comparison with the estimates of the corresponding 1\,Hz data. }
\end{figure*}
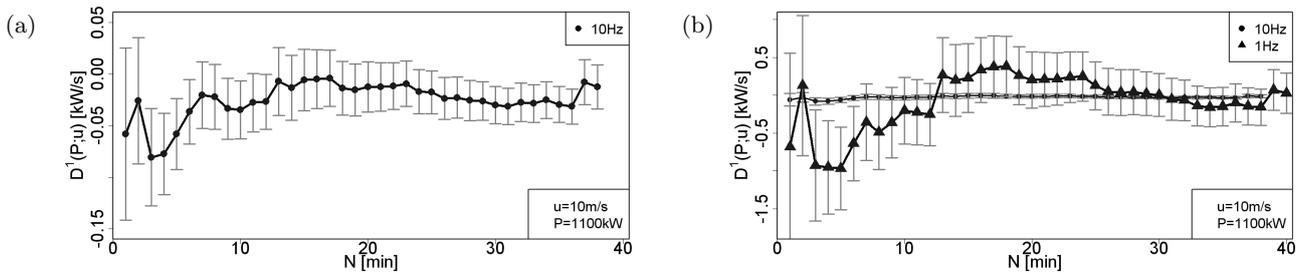

A reliable fixed point can be estimated only if the zero crossing of the $D^{(1)}$ function within the considered wind speed bin remains nearly at the same point for a changing amount of data. 
Thus, for getting a reliable fixed point, converging results for the corresponding $D^{(1)}$ are required.
We display the influence of the amount of data on the accuracy of the estimated drift coefficients by means of one significant coefficient.
Figure~\ref{D1_vgl} shows the evolution of the drift coefficient within the wind speed bin $u$=10$\pm$0.25\,m/s for the corresponding power bin $P$=1100$\pm$20\,kW. 
The wind speed is close to rated wind speed, and the chosen bin is characterized by a high variability in the response dynamics [see Figure~\ref{M1_fit3-8}]. 
In Figure~\ref{D1_vgl}(a), the results for the 10\,Hz data of {\it data5} are shown for an increasing amount of data $N$.
As expected, the error bars decrease as more data points are included in the analysis [see equation~\eqref{error}].
Already after approximately 6 min of data (3600 data points), the estimates converge and deliver similar results. 
However, in comparison with similar estimates for the 1\,Hz data [shown in Figure~\ref{D1_vgl}(b)], the variations of results and error bars are negligible.
The 1\,Hz data clearly show a higher variability in the estimates during the first minutes.
After 10\,min of data (600 data points), the estimates are stabilizing and vary within the error bars around the results of the 10\,Hz data.
Thus, it becomes clear that for our numerical wind turbine model, the results of the 10\,Hz data lead to a much higher accuracy.
This is an important aspect for practical purposes.
Especially for the goal of monitoring the performance of a wind turbine by the Langevin approach, the use of 10\,Hz data will allow to obtain much faster reliable results. 
This issue will be important for our last point of this section on condition monitoring.

Of course, the drift coefficient within one bin can already be calculated from a few single data points, and a first overview of the fixed points can be gained from a relative small amount of data.
Needless to say, for the characterization of a wind turbine's power performance, it is advisable to have more stable results.
In our wind turbine model, a stable result for the drift coefficient is already computable from one single 10\,min time series with a sampling frequency of at least 1\,Hz.
Thus, the local criteria for a reliable slope estimation within one bin can be defined as 10\,min of data or at least 600 data points.
On contrary to the IEC regulation, the amount of data cannot be defined solely in terms of the mean wind speed.
This is due to the fact that the more turbulent the data become, the longer the underlying time series have to be as more power bins are affected per wind speed. 
Each investigated power bin per wind speed has to fulfil the requirement of at least 600 data points to obtain a reliable fixed point.
That means, if four power bins are affected, we need a minimum of 2400 data points. 
Depending on the statistical distribution of the power output for the considered wind speed, the overall needed amount of data for all power bins can vary significantly.
Therefore, the local criteria have to be checked for each bin to evaluate the reliability of the estimated drift coefficients.


\subsection{Site-specific turbulence intensity}\label{ssti}
Former investigations with a very simple numerical model suggest an independence of the Langevin approach from the site-specific turbulence intensity~\cite{Gottschall2008,Boettcher2007}. 
Here, we reinvestigate this proposition with our detailed numerical model of the WP\,1.5 wind turbine by means of two different turbulence intensities: 5\% and 15\%.
As data basis the 10\,Hz data of {\it data5} and {\it data15} are used~[see Table~\ref{tab_seeds}].

\begin{figure*} 
  \begin{minipage}{0.5\textwidth}
    \centering{
      \includegraphics[width=0.95\textwidth]{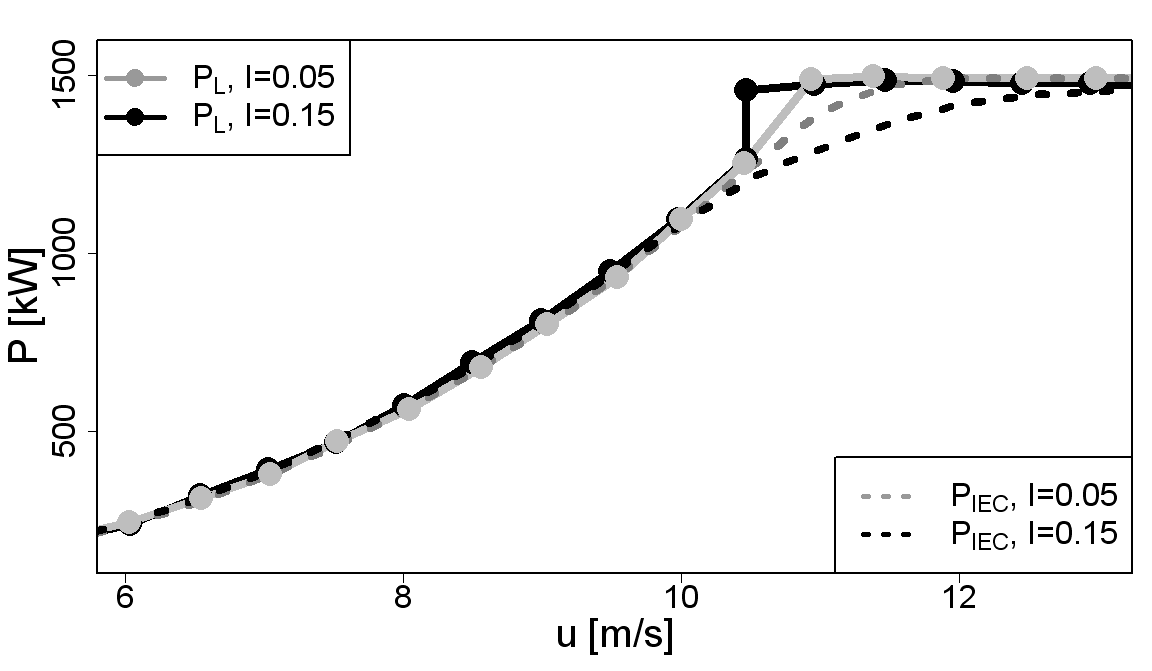}
    }
  \end{minipage}
  \caption[Langevin power curves for 10\,Hz data sets with two different turbulence intensities]{\label{comp15} LPCs for the 10\,Hz data sets of {\it data5} and {\it data15}. The dots represent the fixed points of the LPCs $P_L$, and the corresponding IEC power curves $P_{IEC}$ are given as dashed line. The slope estimation for each bin is carried out with $\tau_{10} \in [0.3,0.8]$.}
\end{figure*}

The LPC estimated from the data with a turbulence intensity of 5\% ($I$=0.05) is already presented in section~\ref{freq}.
Additionally, we determine the LPC on the basis of {\it data15} for a turbulence intensity of 15\% ($I$=0.15).
The fixed points for the more turbulent data could also be calculated with a high certainty as the mean uncertainty is only 1.2\%.
In order to investigate the influence of turbulence on the LPC, we have to compare the results for $I$=0.15 with those from $I$=0.05 presented in Figure~\ref{LPC_beide}(a) and (b).
Figure~\ref{comp15} shows the direct comparison of the two LPCs for all calculated wind speed bins.  
The figure indicates a good correspondence of the LPC  estimated from data with a sampling rate of 10\,Hz. 
Especially for the wind speed bins below 8\,m/s and above 11\,m/s, the fixed points match quite well.
In the range of the wind speed bins from 10 to 11\,m/s, pronounced differences are found, as shown in Figure~\ref{comp15}.
Neglecting this range, the calculated mean root square deviation between the fixed points of both LPCs is 1.4\% over the complete range of wind speeds.  
Please keep in mind that due to the binning, the wind speed may vary a little bit for the compared fixed points. 
Because of the cubic relation of wind speed and power output, a small deviation in the wind speed leads to higher deviation in the power output.  
Thus, the actual mean deviation can be assumed as smaller than 1.4\%.  \\   
For completeness, the common IEC power curves for both turbulence intensities are also shown in Figure~\ref{comp15}.  
As it is well known, the IEC power curve depends on the turbulence intensity and underestimates the power output around rated power~\cite{Boettcher2007}.
The more turbulent the data are, the more pronounced becomes the deviation between the IEC power curves. 
%
%
Comparing the  power curves from the Langevin approach with the IEC power curves, we see that for low wind speeds, both IEC power curves are similar and in good correspondence with the  LPCs.
For higher wind speeds, and especially in the range of transition to full load, the IEC curves deviate significantly from the fixed points, and thus, from the steady states of the wind turbine.
The IEC curves become  site and turbulence dependent. 
The mean root square deviation between the two IEC power curves is 3.5\% over the complete range of wind speeds.
In contrast to this, the LPC only shows an emerging bistability with increasing noise, which is evident for stochastic processes with multistability. 
With increasing noise level, the system will start fluctuating between the different stable fixed point, and thus, in the data analysis the multistability is resolved, as we see in Figure~\ref{comp15}.

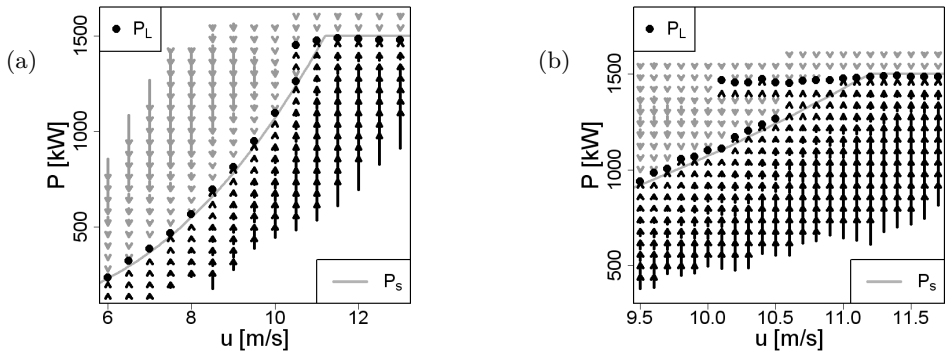
\begin{figure*} 
  \unitlength 1mm
  \centering{
    \begin{minipage}{0.95\textwidth}
      \centering{
        \includegraphics[width=0.3\textwidth]{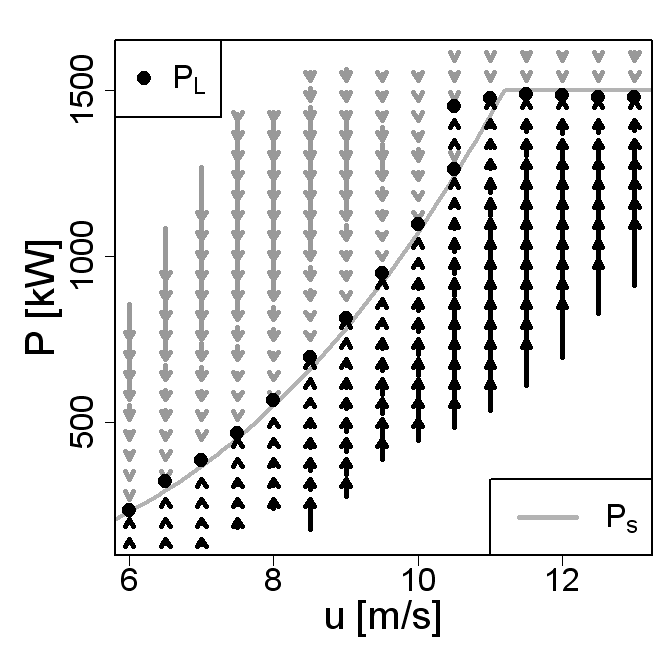}
        \hspace{0.1\textwidth}
        \includegraphics[width=0.3\textwidth]{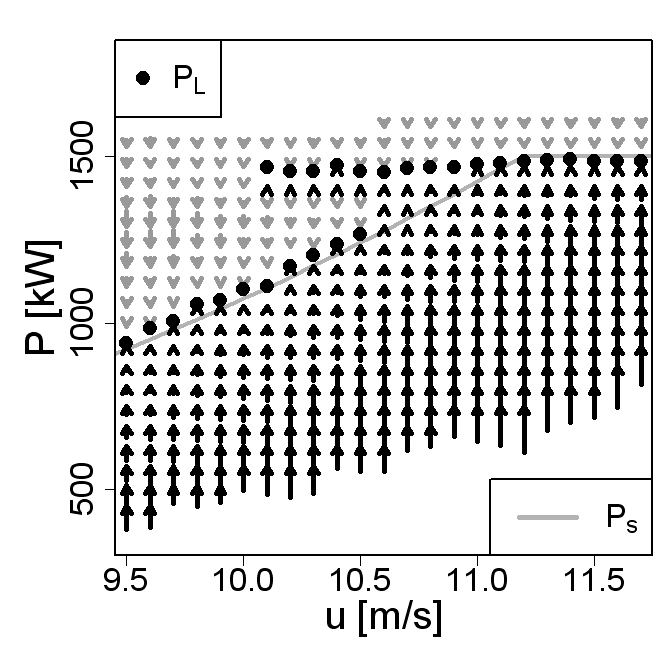}%
        \begin{picture}(0,0)
          \put(-125,40){(a)}
          \put( -55,40){(b)}
        \end{picture}%
      }
    \end{minipage}
  }
  \caption[Drift field for the 10\,Hz data set with 15\% turbulence intensity]{\label{drift15} Drift field for the 10\,Hz data set of {\it data15}. (a)~Complete drift field with a wind speed bin size of 0.5\,m/s and (b)~detail of the drift field with a wind speed bin size of 0.1\,m/s. The solid grey line depicts the steady states $P_s$ of the WP\,1.5 wind turbine and the black dots the fixed points of the conversion process with I=0.15. The slope estimation for each bin is carried out with $\tau_{10} \in [0.3,0.8]$. }
\end{figure*}

In Figure~\ref{drift15}(a), the complete drift field with the estimated fixed points of data set {\it data15} with $I$=0.15 is shown. 
Because of the highly turbulent data, a broad variability of the inflow and power output has developed.
This characteristic is reflected in a more extensive drift field compared with the lower turbulence intensity case [cf. Figure~\ref{LPC_beide}(a)]. 
The bistability of the system at $u$=10.5$\pm$0.25m/s can be identified because of the two fixed points within this wind speed bin.
The arrows illustrate the basins of the two fixed points.
The two black arrows between approximately 1300\,kW and 1400\,kW indicate the change in the dynamic of the conversion process. 
With a smaller wind speed bin size, the onset of the bistability becomes more obvious. 
Figure~\ref{drift15}(b) shows a detail of Figure~\ref{drift15}(a) with a higher resolution of the wind speed (bin size is 0.1m/s).
Obviously,  the dynamics of the wind turbine develops already at 10.1\,m/s two fixed points. 
One of the fixed points coincides with the steady evolution of the fixed point for lower wind speeds, and the other one with the rated power of the turbine.
Such multiple fixed points emerge typically in the range of transition to full load if the control system starts to change its control strategy to rated power~\cite{Milan2010,Milan2009, waechter_dewek}.
It is also interesting to see that the steady states of the uniform flow calculations [grey line in Figure~\ref{drift15}(b)]
do not show the bistability.
Thus, we see that this bistability is caused by the turbulent (noisy) working conditions causing the control system to switch between these fixed points. 
This might be called noise-induced bistability.

\begin{figure*} 
\unitlength 1mm
\centering{
  \begin{minipage}{0.95\textwidth}
   \centering{
    \includegraphics[width=0.28\textwidth]{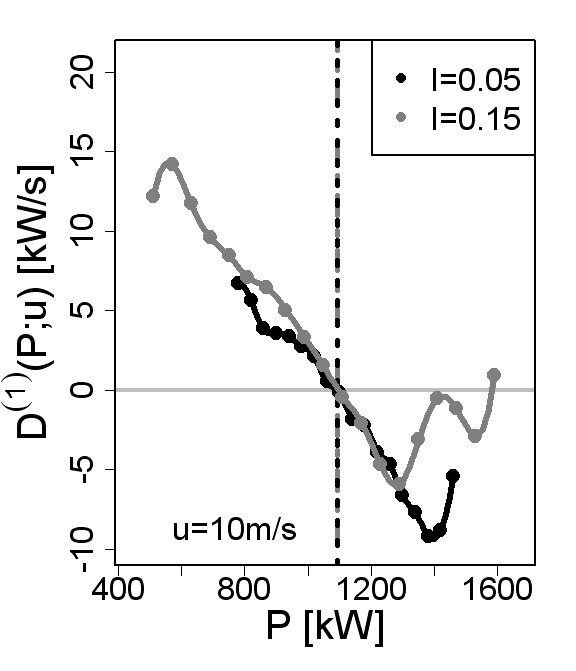}
		 \hspace{0.05\textwidth}
\includegraphics[width=0.28\textwidth]{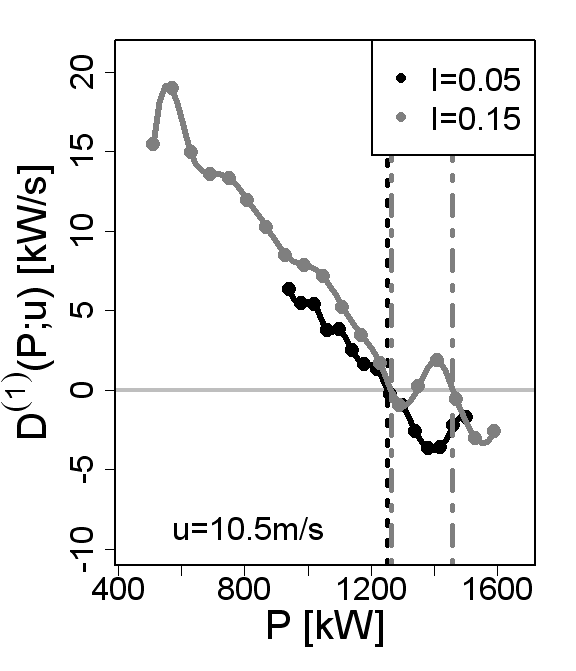}
		 \hspace{0.05\textwidth}
\includegraphics[width=0.28\textwidth]{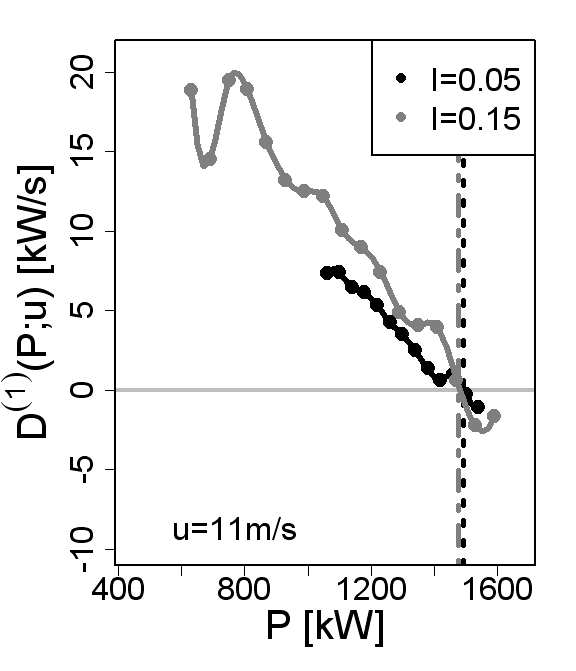}
 	}
  \end{minipage}
  \begin{minipage}{0.95\textwidth}
   \centering{
    \hspace{0.005\textwidth}
    \includegraphics[width=0.28\textwidth]{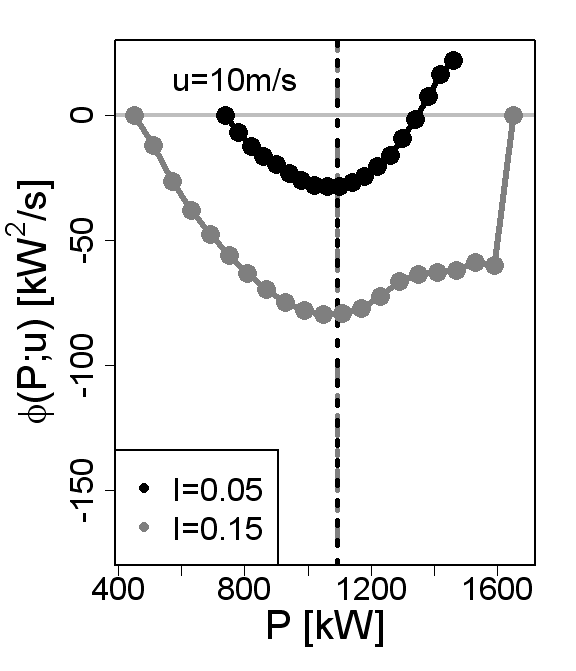}
		 \hspace{0.05\textwidth}
\includegraphics[width=0.28\textwidth]{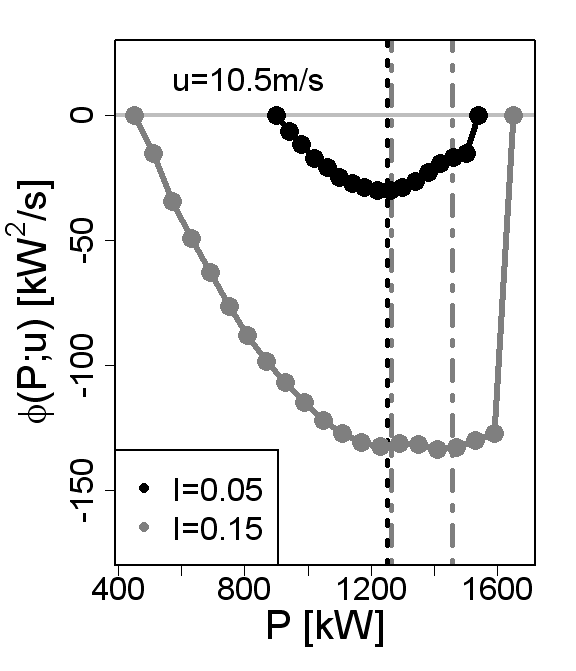}
		 \hspace{0.05\textwidth}
\includegraphics[width=0.28\textwidth]{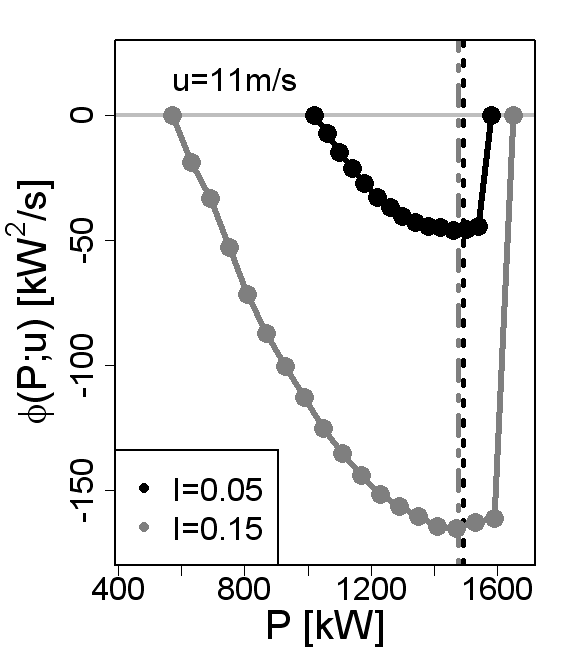}%
\begin{picture}(0,0)
  \put(-167,105){(a)}
  \put(-107,105){(b)}
  \put( -50,105){(c)}
  \put(-167, 48){(d)}
  \put(-107, 48){(e)}
  \put( -50, 48){(f)}
\end{picture}%
 	}
  \end{minipage}
  }
  \caption[Drift functions and potentials for two different turbulence intensities within three wind speed bins]{\label{D1_TI} 
  (a)--(c)~Drift functions $D^{(1)}(P;u)$ and (d)--(f)~potentials $\Phi(P;u)$ for two different turbulence intensities within the wind speed bins 10.0$\pm$0.25m/s, 10.5$\pm$0.25m/s and 11.0$\pm$0.25m/s. The vertical lines depict the stable fixed points of the system for $I$=0.05 (grey dash dotted) and $I$=0.15 (black dotted).}
\end{figure*}
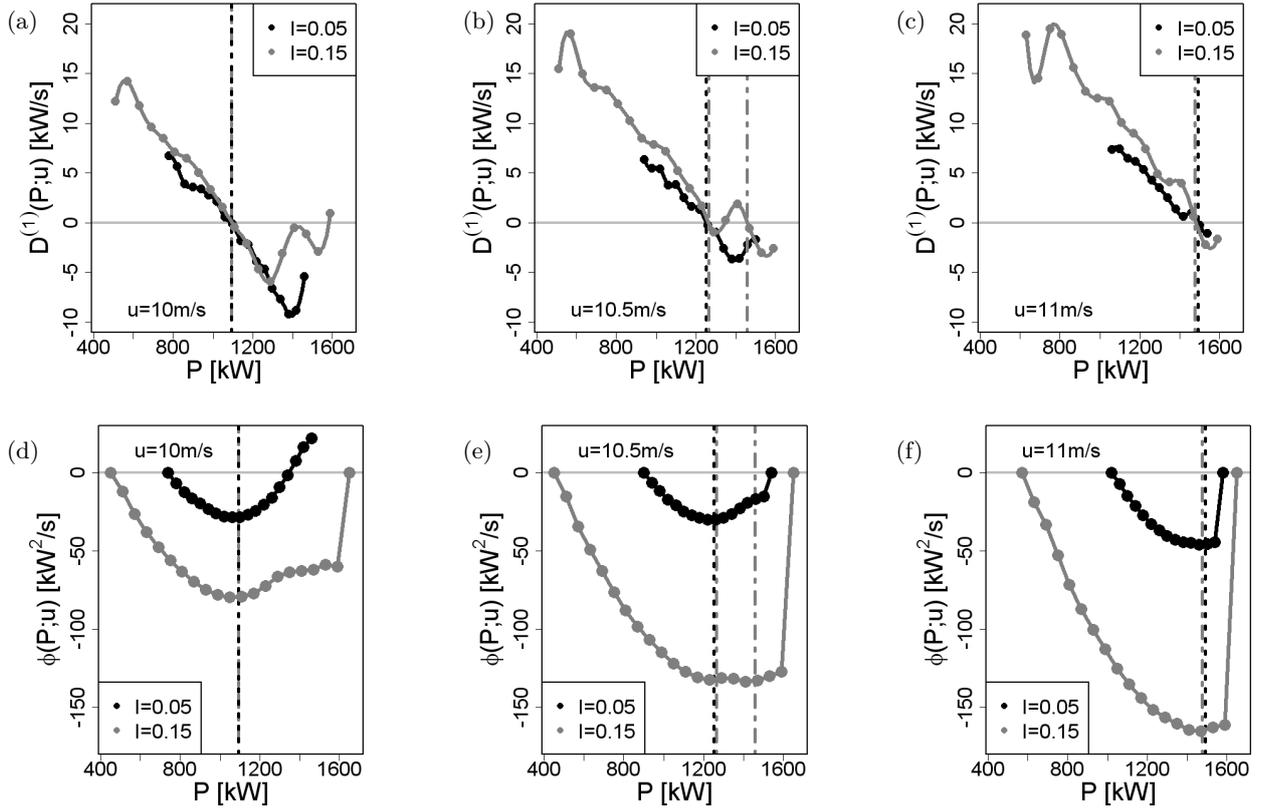

Next, we take a closer look on the differences in the response dynamics for the wind speeds in the range of 10 to 11\,m/s resulting from different turbulence intensities.
Figure~\ref{D1_TI} shows the drift functions and corresponding potentials of the considered wind speed bins. 
As expected, the increased  turbulence intensity causes a wider span of power fluctuation within the wind speed bin [see also Figure~\ref{LPC_beide}(c) in comparison with Figure~\ref{drift15}(a)]. 
Overall, we find that the drift functions for different wind speed bins [Figure~\ref{D1_TI}(a)-(c)] are quite similar whereas the corresponding potentials [Figure~\ref{D1_TI}(d)-(f)] are shifted vertically, which is caused by the integration over a larger range in the power output $P$ and has no physically meaning.
In addition, the fixed points (zero crossing of the drift function and minimum for the potential) are marked by vertical lines.
We see how the fixed point shifts from about 1100\,kW for $u$=10\,m/s to 1400\,kW as a function of the wind speed $u$.
Already for $u$=10\,m/s [see Figure~\ref{D1_TI}(a) and (d)], we find that the drift function $D^{(1)}$ for $I$=0.15 becomes almost zero close to 1350\,kW, without crossing the $P$-axis.
In the corresponding potential, we find a constant level, indicating a metastable situation.
This metastability evolves for high wind speed bins into a stable fixed point; i.e. the drift function crosses the $P$-axis with a negative slope, and the potential $\Phi$ shows a clear minimum.
Summarizing these findings in terms of the potentials, we could unfold with our Langevin method the transition of the system from one stable minimum in the potential, over a double well potential with two minima, to the state where finally solely the new fixed point and one minimum in the potential are left.
Most interesting, this ''phase transition''  seems to be continuously evolving and to be forecasted by the deformation of the potential.\\
The comparison of the results for different turbulence intensities shows in detail some significant differences.
This is in compliance with the previously mentioned statement of noise-induced dynamics.
As the system is reacting by its control system on different wind situations, we find in this transition region to rated power turbulence dependent and, thus, site dependent dynamics of the power conversion. 
Once the new stable fixed point is established [see Figure~\ref{D1_TI}(c) and (f)], the results become again site independent.
We can conclude that for the data set of higher turbulence intensity, the bistability of the system becomes more pronounced.
Comparable results are also found in case of the 1\,Hz and the LIDAR-like data, which confirms that more information about the different control strategies are gained because of higher noise.
With a lower turbulence of 5\%, the wind turbine remains longer and more often in the lower power state, and the higher second fixed point is not evolving.

The presented results of the detailed numerical wind turbine model affirm that the LPC is independent of the site-specific turbulence characteristics. 
The position of the fixed points are not affected by the turbulence, whereas for a higher turbulence intensity, the fixed point at rated power can be detected already for lower wind speed values.
The LPC matches the steady states of the wind turbine for both investigated turbulence intensities.
Because of the higher variety of wind and power fluctuations, more turbulent data can give additional information, especially about the response dynamics in the transition range  to full load.
The corresponding drift functions and potentials provide a deeper insight into the control strategy of the wind turbine.
Fixed points as well as critical points of the conversion process can be identified.
Hence, the estimated LPC gives a compact representation of the power performance of the entire WEC.


\subsection{Condition monitoring}
Malfunctions or faulty inflow conditions spoil the conversion dynamics of a wind turbine and lead to modifications in the relation of wind speed and power output.
It has been demonstrated that the LPC represents the power performance of the entire WEC; thus, it should be possible to detect deviations from the normal operation in detail.  
Therefore, it is an obvious approach to transfer the Langevin method to the condition monitoring of wind turbines. 
The idea of power performance monitoring is to compute an initial LPC under normal operating conditions serving as a reference power curve. 
Deviations from this reference denote a failure in the extraction process~\cite{Milan2010}.
In this section, we show that the method works quite well and detects failures in the conversion process very fast.
Two different failure scenarios are presented: a malfunction due to a pitch failure, which leads to an increased power production, and power loss due to a yaw error (faulty inflow condition). 
The first reference LPC is data set {\it data5}, and the second is {\it data15}. 
In order to generate a power curve containing the malfunction, the simulated failure files are added to the reference files, and the cumulated LPC~$P_{LF}(u)$ is calculated with the same settings as the reference LPC~$P_L$. 
Regarding the analysis, it makes no difference if the error occurs continuously or intermittently during the time of operation.
All deviations from the reference LPC are driven by the joined failure durations in the cumulated LPC.
We will see that each failure has an indicating fingerprint and can be identified by the described procedure.
Furthermore it becomes clear that the method is not restricted to a certain turbulence intensity.
In the following figures, the reference LPC is always displayed in black, and the cumulated LPC containing the failure in grey.

\subsubsection{Pitch failure}
In this first example, we show an error of the pitch system during operation, which leads to an increased power output above rated wind speed. 
The LPC~$P_L(u)$ with the turbulence intensity $I$=0.05 and 10\,Hz sampling rate from Figure~\ref{LPC_beide}(a) is taken as reference.
This low turbulence intensity is characteristic for offshore or near-shore sites.
For each wind speed, data files of 10 and 60\,min duration have been generated containing a pitch error.
This corresponds to a failure ratio of 3.2\% and 19.2\% in the merged data sets.
The pitch error is reflected by a fixed pitch angle for all three blades.
More precisely, the same initial angle applies for each wind speed as in the partial load range. 
Thus, the WEC is not regulated to rated power, and the power output is continuously increasing.
Under normal operating conditions such an extreme increase of the power output would be avoided with a shut down initiated by the control or safety system.  
However, in particular for an intermittent occurrence of the failure for the power output around rated wind speed, such a pitch failure would not automatically lead to a shut down as  the operating parameters of the wind turbine are within their normal limits. 
Furthermore, the occasional exceedance of operating limits for short time periods, like several seconds, will not necessarily cause significant changes in the monitored mean values of the operating parameters.

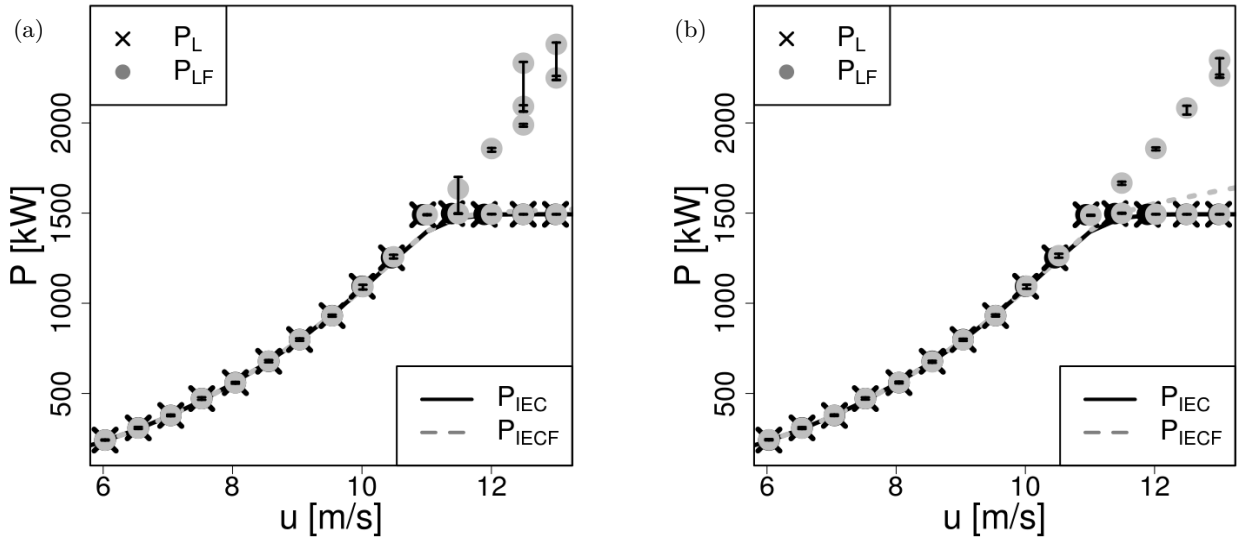
\begin{figure*} 
  \unitlength 1mm
  \centering
  \begin{minipage}{0.95\textwidth}
    \centering
    \includegraphics[width=0.45\textwidth]{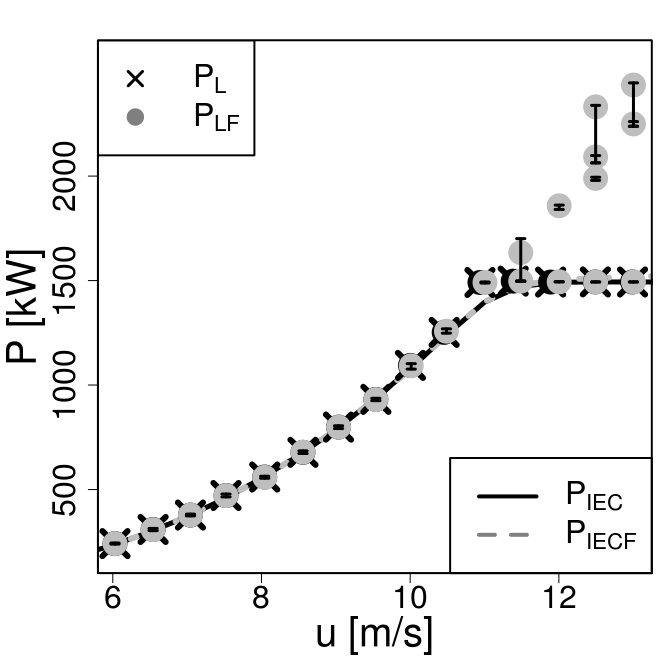}
    \hspace{0.05\textwidth}
    \includegraphics[width=0.45\textwidth]{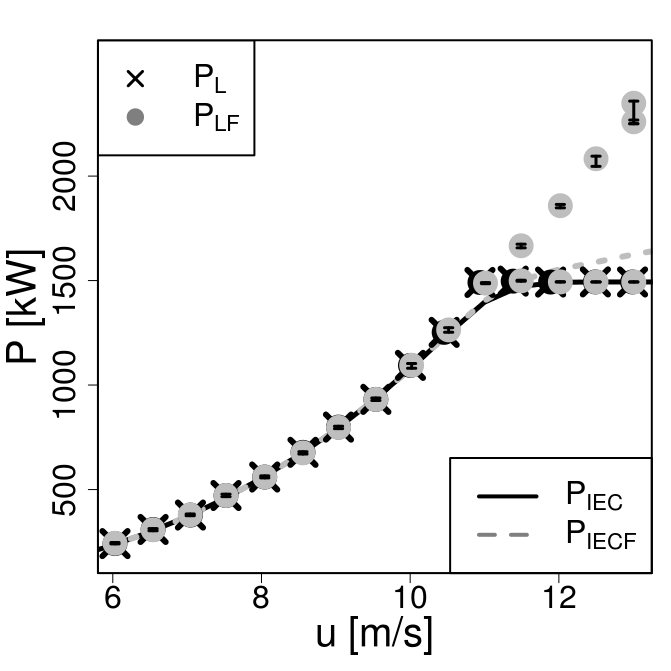}%
    \begin{picture}(0,0)
      \put(-163, 68){(a)}
      \put( -76, 68){(b)}
    \end{picture}%
  \end{minipage}
  \caption[Comparison of the reference LPC of {\it data5} with the LPC of the partly disturbed conversion process due to an added pitch failure]{\label{LPC_addStall-Err} Comparison of the reference LPC~$P_L$ (black crosses) of {\it data5} with the LPC~$P_{LF}$ (grey dots) of the partly disturbed conversion process due to an added pitch failure lasting (a)~10\,min  and (b)~60\,min, that is 3.2\% or 19.2\% of the running time per wind speed bin, respectively. Error bars are shown for $P_{LF}$. 
The corresponding IEC power curves are shown for both cases, as grey-dashed line for the failure and as black solid line for the reference case.}
\end{figure*}

Figure~\ref{LPC_addStall-Err} shows the LPC (grey dots with error bars) including different pitch failure durations together with the reference LPC (black crosses).
In Figure~\ref{LPC_addStall-Err}(a), the results for a 10\,min failure period per wind speed bin are shown; in Figure~\ref{LPC_addStall-Err}(b), those for six times longer failure periods are shown.
Additionally, the results for the reference IEC power curve (black line) and the one including the failure (grey dashed line) are displayed.
Because of the short-time occurrence of the pitch failure, there is merely a marginal increase in the IEC curve for the wind speeds above 11\,m/s  [Figure~\ref{LPC_addStall-Err}(a)].
This increase becomes more pronounced for a longer duration of the pitch failure.
However, as the IEC power curve is an average of all values for one wind speed bin, its shape is changing slowly.
Just in the case the new dynamics has a sufficient portion in the data accumulation, the mean value varies significantly [Figure~\ref{LPC_addStall-Err}(b)]. 
In contrast to this, the LPC develops immediately new fixed points above rated wind speed.
The doubled fixed points quantify the exact deviation from the reference case and are not affected by any averaging effects.
For a longer duration of the failure, the additional fixed points of the LPC become more definite, and the error bars decrease.
Below rated power, the response dynamics of the added failure files coincides as expected with the reference case, and the fixed points of $P_{LF}(u)$ remain unchanged in the range from 6 to 10\,m/s.
%

\begin{figure*} 
  \unitlength 1mm
  \centering{
    \begin{minipage}{0.95\textwidth}
      \centering{
        \includegraphics[width=0.27\textwidth]{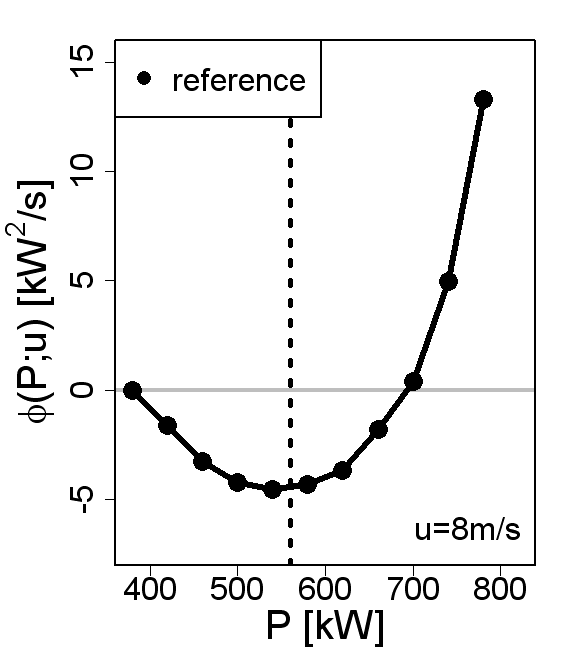}
        \hspace{0.06\textwidth}
        \includegraphics[width=0.27\textwidth]{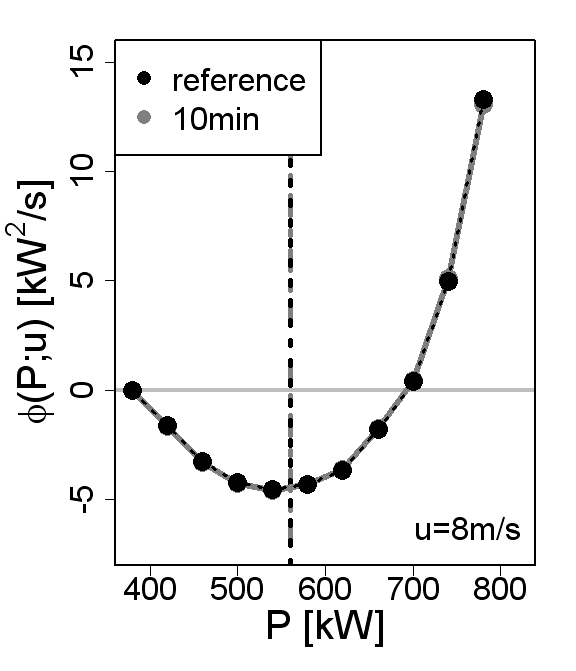}
        \hspace{0.06\textwidth}
        \includegraphics[width=0.27\textwidth]{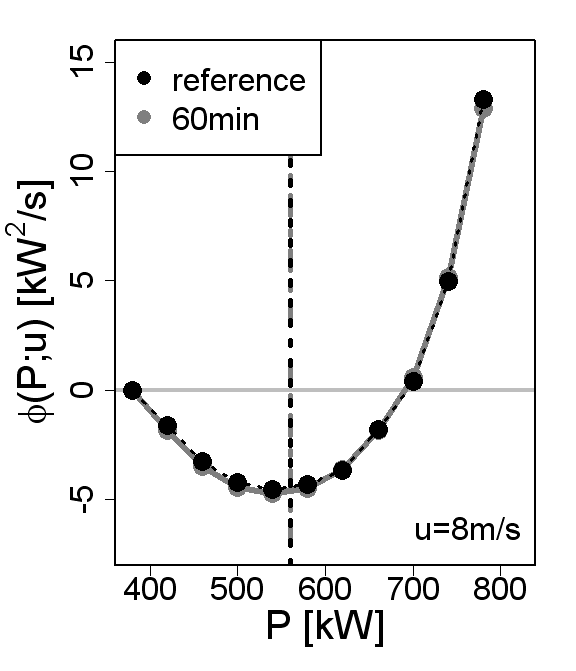}
      }
    \end{minipage}
    \begin{minipage}{0.95\textwidth}
      \centering{
        \hspace{0.01\textwidth}
        \includegraphics[width=0.27\textwidth]{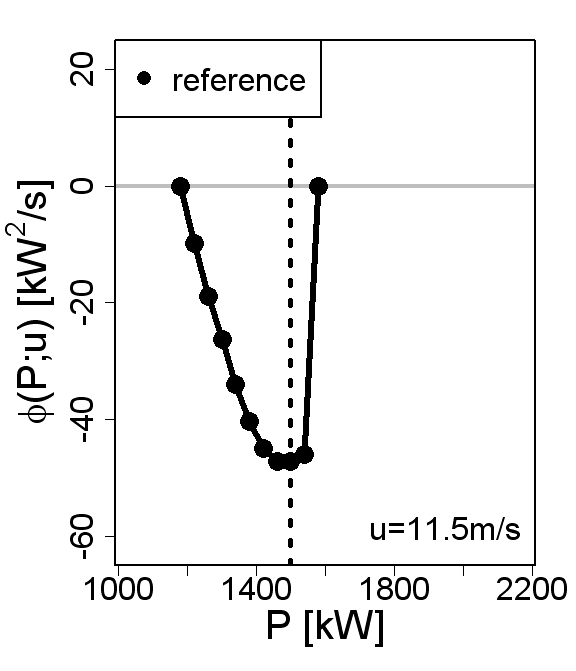}
        \hspace{0.06\textwidth}
        \includegraphics[width=0.27\textwidth]{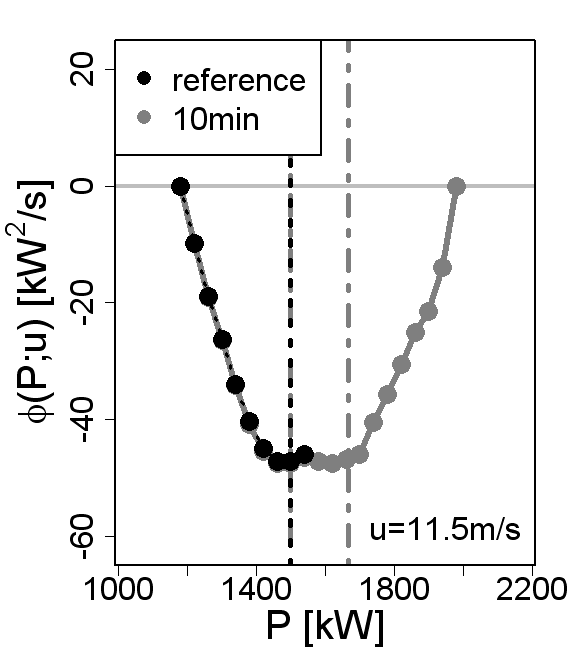}
        \hspace{0.06\textwidth}
        \includegraphics[width=0.27\textwidth]{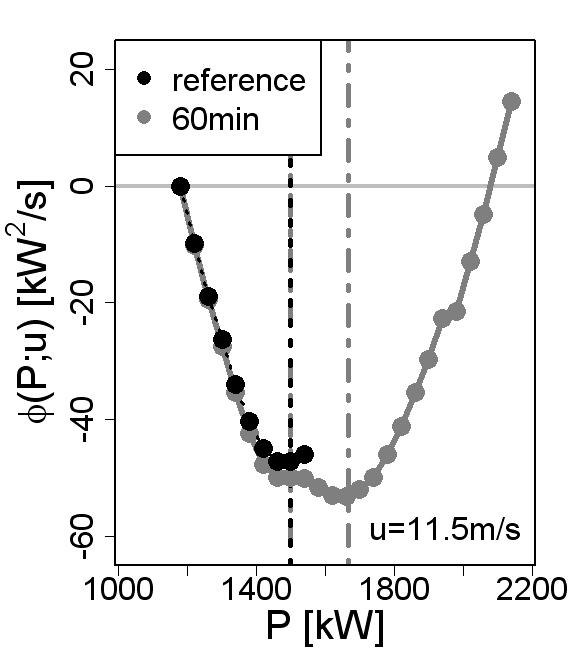}%
        \begin{picture}(0,0)
          \put(-167,102){(a.1)}
          \put(-107,102){(a.2)}
          \put( -50,102){(a.3)}
          \put(-167, 48){(b.1)}
          \put(-107, 48){(b.2)}
          \put( -50, 48){(b,3)}
        \end{picture}%
      }
    \end{minipage}
  }
  \caption[Potentials within different wind speed bins for the reference LPC in comparison with a failure LPC including a pitch failure]{\label{Potential_addStall} 
  Potentials $\Phi(P;u)$ within the wind speed bins (a)~$u$=8.0$\pm$0.25\,m/s and (b)~$u$=11.5$\pm$0.25\,m/s for (.1)~the reference case~(black), (.2)~the reference case together with 10\,min pitch error per wind speed bin (grey) and (.3)~60\,min pitch error per wind speed bin (grey). The vertical lines are displaying the position of the fixed points at (a)~560\,kW and (b): (b.1)~1498\,kW; (b.2)~1498 and 1634\,kW; (b.3)~1498 and 1667\,kW.}
\end{figure*}
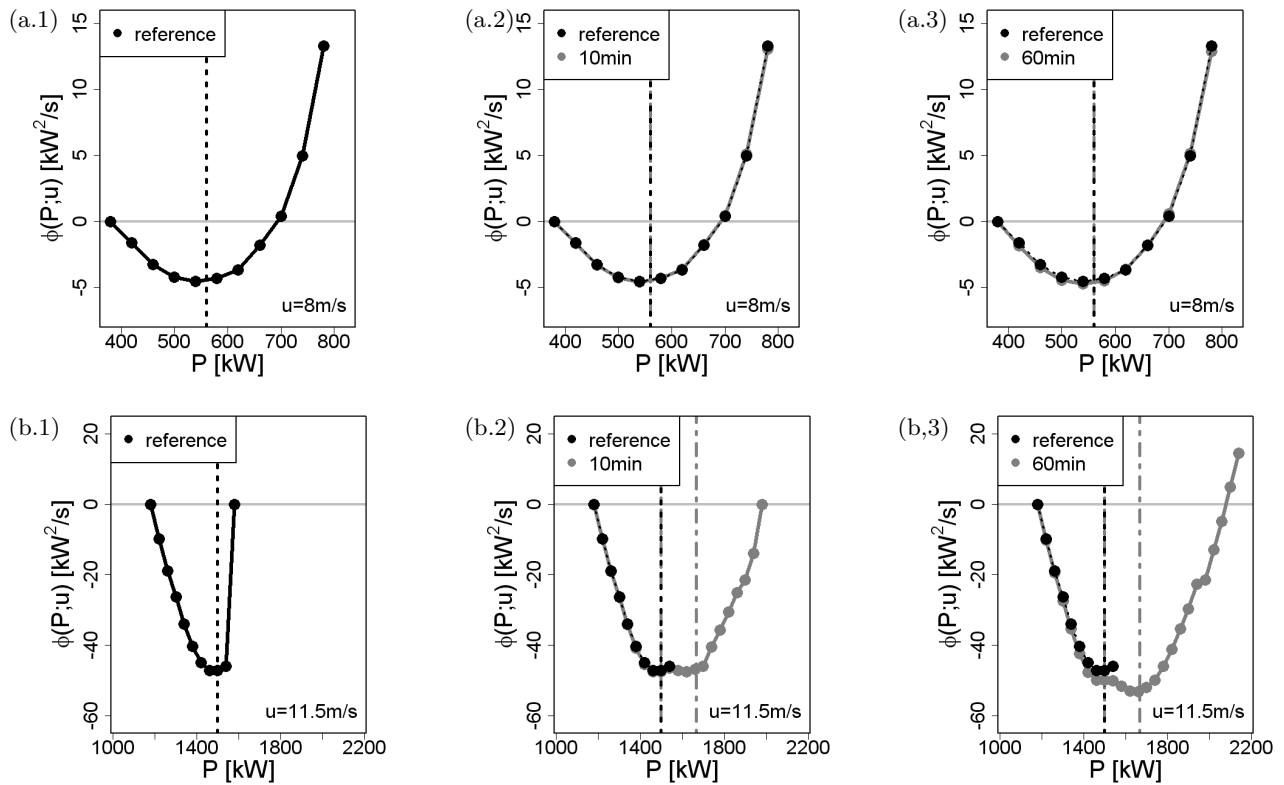

Next, we consider the corresponding potentials. 
Below rated power, the potentials remain unchanged~[Figure~\ref{Potential_addStall}(a)].
Above rated power, the potentials deviate from the reference and become bistable.
As shown in Figure~\ref{Potential_addStall}(b), in addition to the previous minimum, a second one develops immediately after the occurrence of the failure.
In each chart, the black dots mark the undisturbed reference potential, and the grey dots the potential including the failure. 
In (b.1) only the potential~$\Phi(P)$ of the reference case is given (black dots). 
The other two subfigures show the reference potential together with the newly developed potentials (grey dots). 
The reference potential  has one minimum at 1498\,kW, which is the stable fixed point of this wind speed bin~[Figure~\ref{LPC_addStall-Err}(a)].
With 10\,min pitch failure, the potential becomes bistable. 
An additional minimum occurs at 1634\,kW with a small barrier between both stable states, where an unstable fixed point is located. 
For 60\,min pitch failure, the additional minimum and the accompanied bistability become more pronounced.
The system has now two separated stable fixed points at 1499 and 1667\,kW.
With an increasing amount of data, i.e. longer duration of the failure, the additional valley becomes deeper, and the position of the fixed point will slightly change.
It is important to note that the additional fixed point can be identified clearly for both cases.
The depth of the additional minimum may be taken as a signature of the importance.


\subsubsection{Yaw error}
As second example, we consider a yaw error in the range of partial load.
In cold climate or because of mechanical failures, it may occur that the measurement equipment of the wind turbine does not operate faultlessly.
In this case, a yaw error can occur because of a malfunction of the wind direction measurement device.
Here, we simulate a yaw error of 43\degree in the range of 8\,m/s, whereas the measurement of the magnitude of wind speed is working properly.
With such an inflow, the machine is not operating under optimal aerodynamic conditions, and the power output will decrease. 
In the ideal uniform case of our wind turbine model, the simulated yaw error causes a change in the power output from 550.7\,kW to 220.7\,kW for an inflow of 8\,m/s.
Thus, the power is decreasing by around 60\%.
The LPC $P_L(u)$ with a turbulence intensity of $I$=0.15 and 10\,Hz sampling rate is taken as reference [see Figure~\ref{drift15}(b)].
This turbulence intensity is a common value for onshore sites with moderate terrain.
The analysis is carried out similar to the former example with the difference that we consider only additional yaw failure files in the region around 8\,m/s.
The failure files were generated for the mean wind speeds $\bar u$=7.5, 8.0 and 8.5m/s.
As in the example before, files with failure  of 10 and 90\,min durations are generated for each wind speed for our LPC analysis.
This corresponds to a failure ratio of 1.6\% and 14\% in the merged data sets.

\begin{figure*} 
  \unitlength 1mm
  \centering{
    \begin{minipage}{0.95\textwidth}
      \centering{
	\includegraphics[width=0.45\textwidth]{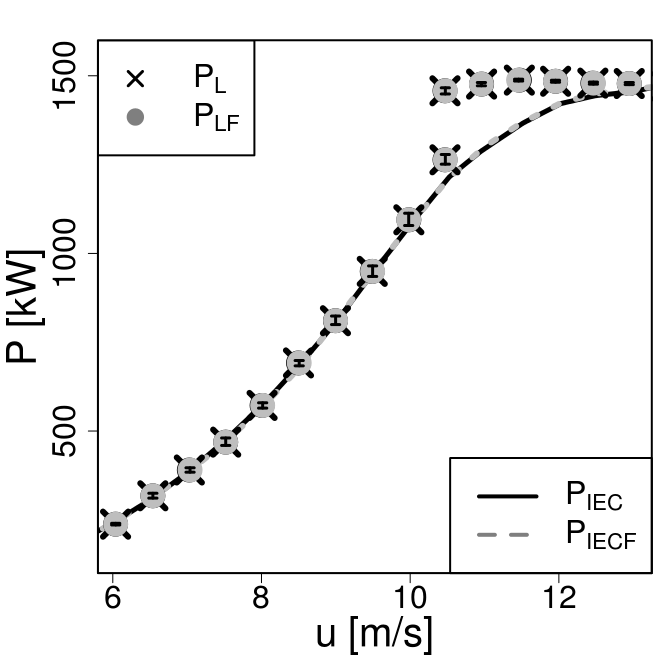}
        \hspace{0.05\textwidth}
        \includegraphics[width=0.45\textwidth]{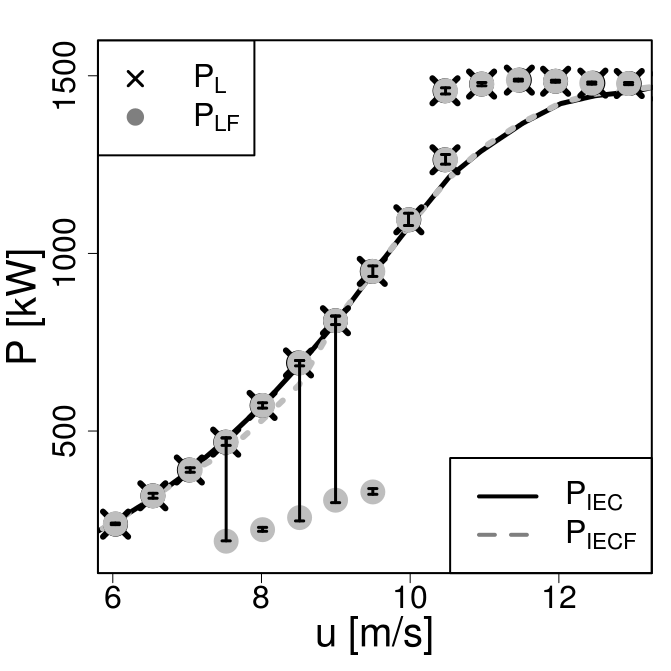}%
        \begin{picture}(0,0)
          \put(-164, 68){(a)}
          \put( -76, 68){(b)}
        \end{picture}%
      }
    \end{minipage}
  }
  \caption[Comparison of the reference LPC of {\it data15} with the LPC of the partly disturbed conversion process due to a yaw error]{\label{LPC_addYaw-Err} 
  Comparison of the reference LPC $P_L$ (black dots) of {\it data15} with the LPC $P_{LF}$ (grey dots) of the partly disturbed conversion process due to a yaw error lasting (a)~10\,min and (b)~90\,min per mean wind speed $\bar u$=7.5, 8.0, 8.5\,m/s, respectively. 
The displayed error bars are given for $P_{LF}$. 
The IEC power curves are calculated for both cases and shown as lines: dashed grey for the failure $P_{IECF}$ and solid black for the reference case $P_{IEC}$.}
\end{figure*}
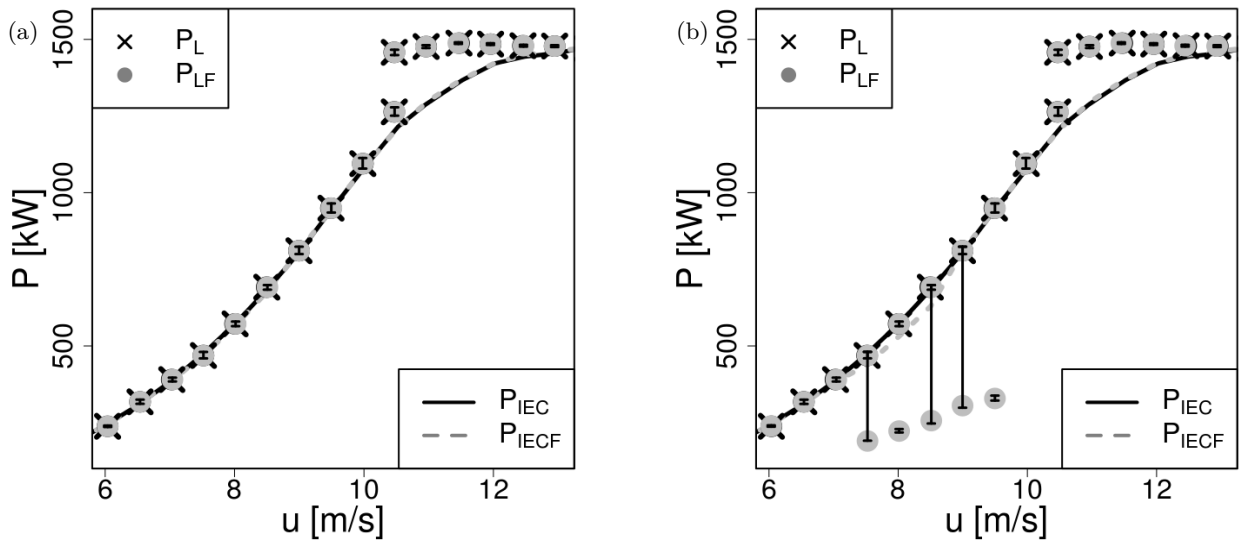

Figure~\ref{LPC_addYaw-Err} shows the LPCs of the reference case {\it data15} (black crosses) together with  the merged data sets of the yaw failure LPC $P_{LF}$ (grey dots).
It can be seen clearly in Figure~\ref{LPC_addYaw-Err}(a) that the IEC power curves are not affected by the short-time malfunction.
Thus, only one 10\,min failure per mean wind speed bin cannot be grasped by  the IEC curve.
For a longer duration of the error, the IEC power curve is slightly reduced in the range of the corresponding wind speed bins [Figure~\ref{LPC_addYaw-Err}(b)].
Neither all affected wind speeds nor the size of the deviations can be analysed correctly by the IEC power curves.
In this respect again, the Langevin approach shows a different behaviour in comparison to the IEC curve.
During the short failure duration, the fixed points of the LPC $P_{LF}$ remain unaffected. 
After a longer duration of the failure, additional fixed points emerge at the failure-related wind speed bins.

\begin{figure*} 
  \unitlength 1mm
  \centering{
    \begin{minipage}{0.95\textwidth}
      \centering{
        \includegraphics[width=0.27\textwidth]{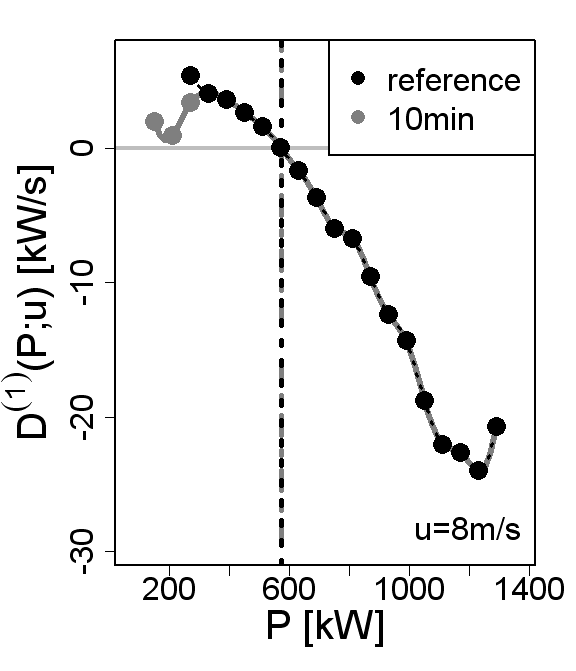}
        \hspace{0.06\textwidth}
        \includegraphics[width=0.27\textwidth]{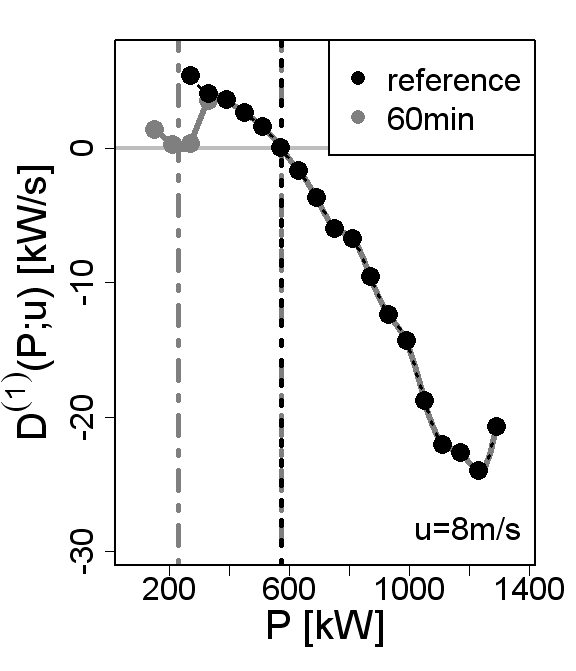}
        \hspace{0.06\textwidth}
        \includegraphics[width=0.27\textwidth]{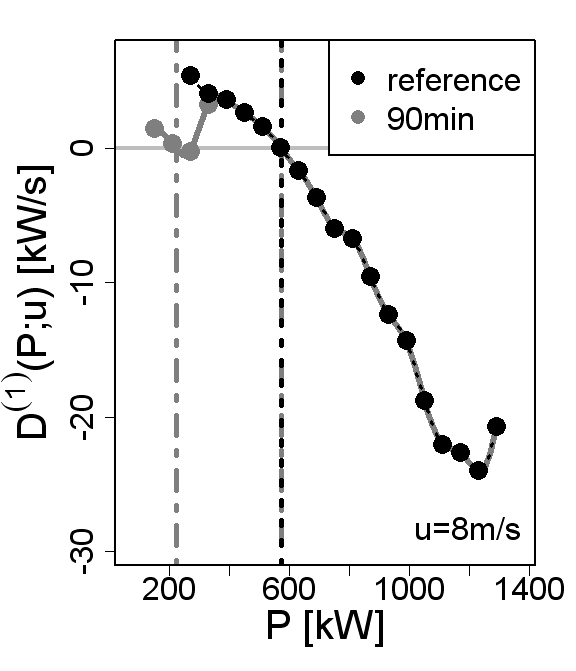}
      }
    \end{minipage}
    \begin{minipage}{0.95\textwidth}
      \centering{
        \hspace{0.01\textwidth}
        \includegraphics[width=0.27\textwidth]{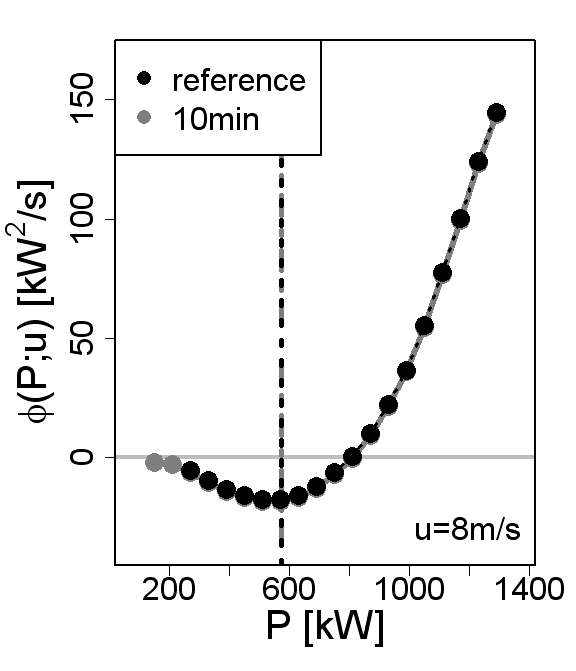}
        \hspace{0.06\textwidth}
        \includegraphics[width=0.27\textwidth]{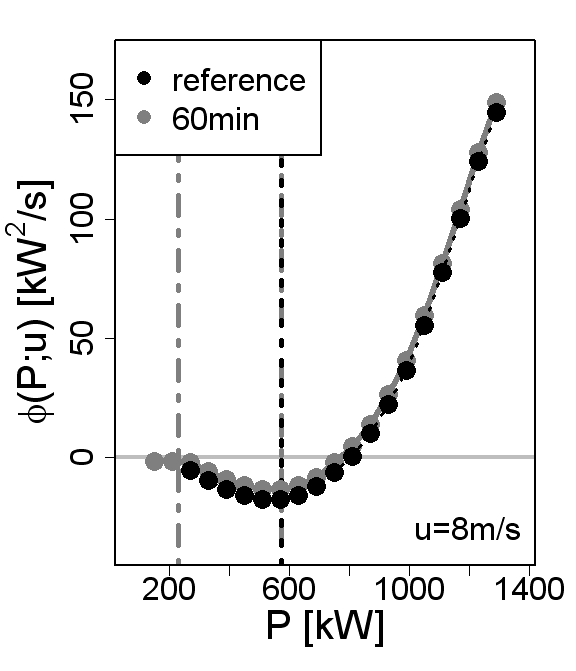}
        \hspace{0.06\textwidth}
        \includegraphics[width=0.27\textwidth]{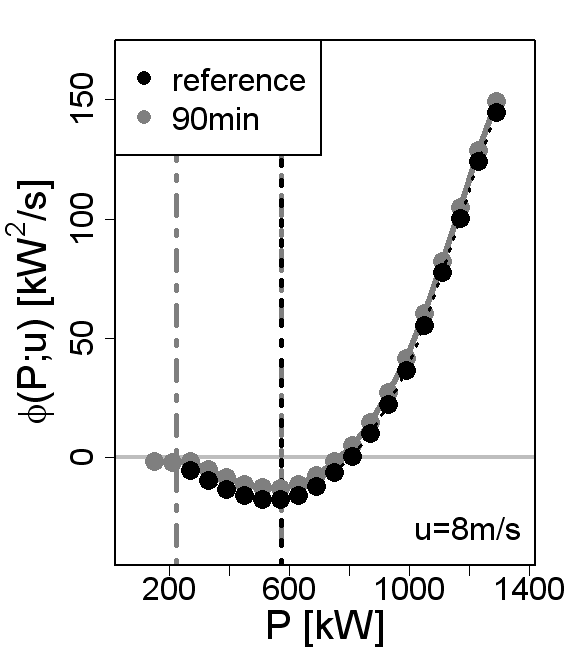}%
        \begin{picture}(0,0)
          \put(-167,102){(a)}
          \put(-107,102){(b)}
          \put( -50,102){(c)}
          \put(-167, 48){(d)}
          \put(-107, 48){(e)}
          \put( -50, 48){(f)}
        \end{picture}%
      }
    \end{minipage}
  }
  \caption[Drift fields and potentials within different wind speed bins for the reference LPC in comparison with a failure LPC including a yaw error]{\label{LPC_addYaw-Phi} (a)~Drift functions $D^{(1)}(P;u)$ and (b)~potentials $\Phi(P;u)$ within the wind speed bin 8.0$\pm$0.25\,m/s for the reference case~(black) together with (.1)~10\,min, (.2)~60\,min and (.3)~90\,min yaw errors (grey) within the considered wind speed bin. The vertical lines show the position of the fixed points at (b.1)~572.1\,kW; (b.2)~230.5 and 572.1\,kW; (b.3)~222.5 and 572.1\,kW.}
\end{figure*}
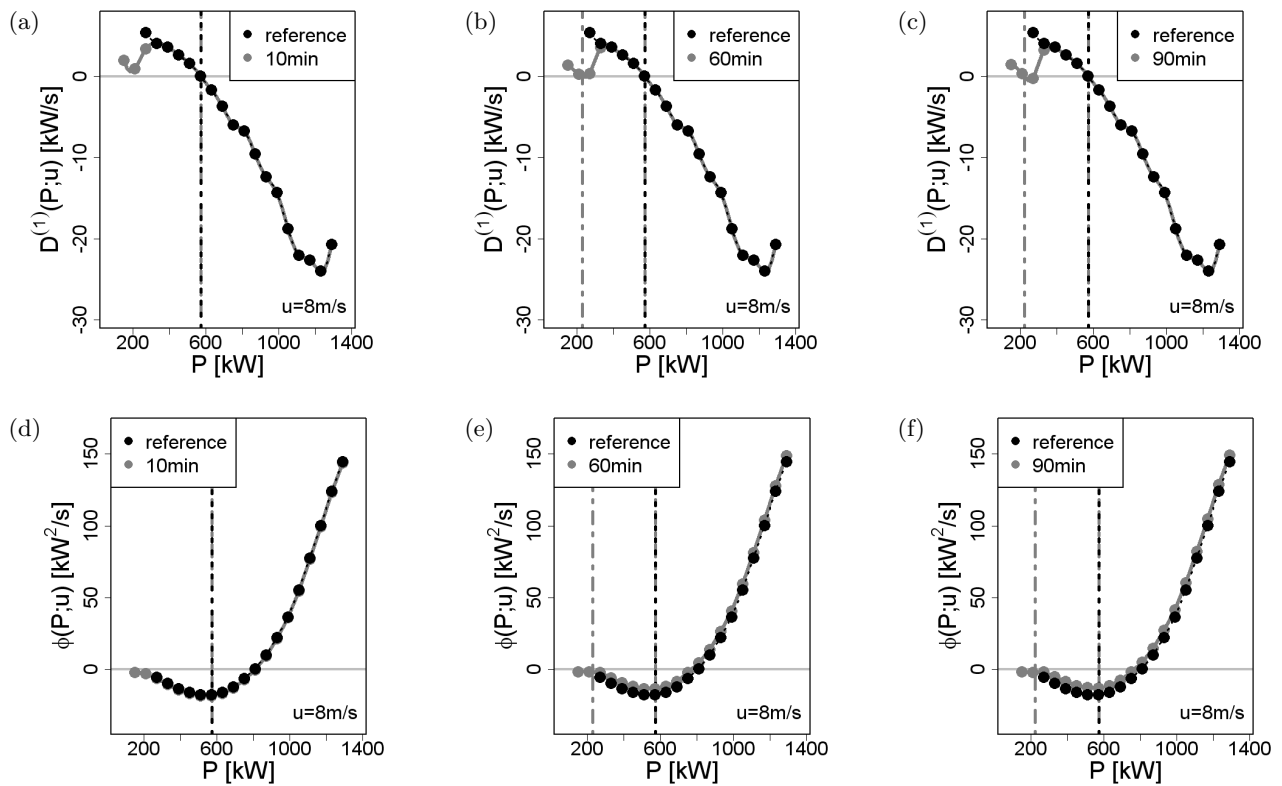

In order to obtain a better understanding why the LPC curves do not always show indications of the failures, we analyse the findings of the LPC in more detail by the use of the drift functions. 
In Figure~\ref{LPC_addYaw-Phi}, the evolution of the drift functions and the potentials are displayed for the wind speed bin 8\,$\pm$0.25\,m/s. 
As we have not found clear signatures of the failures in the LPC, we analyse here the drift functions and the potentials for failure durations of 10, 60 and, additionally, 90\,min.
Whereas the potentials show only minor changes in their shape for these failure cases,  we see pronounced modifications in the drift functions. 
Already for the 10\,min failure case, we find a significant change in the drift function for values below 400\,kW. 
This change indicates the tendency to build up a new crossing of the drift function with the zero line; i.e. the system shows the tendency to build up a new fixed point.
As the failure time increases, the emergence of the new stable fixed point at about 220\,kW is confirmed. 
A closer look on the potential also confirms the multistability for the case of longer failure durations.
Additionally, we have examined failure cases with a smaller error angle in the yaw system. 
In such situations, double fixed points needed a longer time period to emerge.
Nevertheless, in any case, deformations in the drift functions are found at an early stage and, however, before pronounced double fixed points are displayed in the LPC.

Based on the results and discussions of the previous sections, it becomes clear that it is important to have a good understanding of the quality of the determined coefficients and their interpretations. 
We find here that the drift functions are the important quantity to look at,to see how a yaw error slowly emerges.
For our case, a malfunction is indicated after 10\,min and can be detected clearly after 90\,min.


\section{Conclusions}
The Langevin approach estimates the stationary states or fixed points by extracting the actual deterministic dynamics of the wind turbine. 
Stochastic influences are handled as noise and separated from this information. 
Thus, we obtain a power characteristic that is representative for a specific wind turbine but independent of site-specific parameters.
Beyond the stationary states of the process, i.e. the LPC, it provides the complete drift characteristic and therewith additional relevant information about the  short-time dynamics of the WEC.
Just for this reason, the dynamical method is of another type than the IEC standard and cannot replace it without further adjustment. 
Instead of mean values, we estimate fixed points, which has essential implications on the application of the resulting power curves~\cite{Gottschall2008}.

In this paper, we summarized the Langevin approach and gave an introduction of how to obtain a reliable LPC from numerical wind turbines models. 
We found out that high frequent wind speed and power output data with a sampling rate of at least 1\,Hz are in general usable for the estimation of an LPC.
Dependent on the intrinsic response dynamics of the considered wind turbine, it is  necessary to use a higher sampling rate to gain detailed information.
For our wind turbine model, the use of 10\,Hz data is of clear advantage. 
For a different type of wind turbine, corresponding analyses have to be performed in order to find the appropriate sampling rate. 
In general, faster control systems require faster sampling rates. 

At present, it is customary to arrange the wind speed and power bins side to side over the whole range of data sets.
With this procedure, an estimated fixed point is reliable if the calculated drift function is based on at least 600 data points within each power bin. 
That means a higher quantity  of simulated data sets is required for more turbulent data as the power output is more spread and the drift field is broader distributed.
A possibility  for the reduction of the total quantity of data is the use of overlapping bins.
The  precise effect on the data quantity is the basis of further investigations.

Here, we followed the common procedure and generated a sufficient quantity of data according to the reliability requirement.
Therewith, we derived successfully  the power characteristics of a numerical 1.5\,MW wind turbine model, following the IEC standard as well as the novel Langevin approach.
The estimated LPC gives a compact representation of the power performance of the entire WEC.
We showed that the estimated fixed points of the LPC reflect the steady states of the wind turbine and that they can give detailed information about the control strategy of the wind turbine.
This is an important additional information to the IEC power curve. 
We have shown that the data for meaningful LPCs of field measurements can be received from ultrasonic or cup anemometers as long as the sampling rate is sufficient.
In general, LIDAR data are also usable, but because of the spatial averaging during the measurements, the estimated LPCs could only display an averaged power output.
The respective quality of the LPC depends on the specific averaging method of the LIDAR and the response time of the considered control system. 
With our LIDAR-like inflow, it is possible to display the steady states of the WP\,1.5 wind turbine and to estimate the fixed points without the strong underestimation of the power output in the range of rated power, which is typical for the IEC method.

The IEC power curve is strongly affected by the wind conditions during the measurement. 
Especially for a high turbulence intensity, the results deviate significantly from the steady states of the wind turbine.
Observing such deviations in the IEC power curve of a real turbine may raise the question of the causes.
Is this a malfunction of the turbine or just a consequence of the changed turbulence intensity?
 For estimated LPCs, a clear answer can be obtained.
Our investigations with the detailed numerical wind turbine model affirm the independence of the LPC on the site-specific turbulence characteristics. 
If the LPC matches the steady states of the wind turbine independently from the turbulence intensities, we can conclude that 
turbine is still operating correctly. 
It is worth mentioning that the conditioned Langevin method even works better with more noise as we have more dynamics and can obtain more information about the system.

Once the characteristics of a wind turbine are described by the LPC, deviations of the wind turbines performance are easily detectable.
The LPC is able to detect anomalies in the response dynamics over the full range of wind speeds, no matter if the wind turbine is operating under partial or full load. 
In most cases, a deviation from the normal operating condition of a wind turbine will cause immediately severe changes in the affected drift function, what is also seen in the deformation of the corresponding potentials.
After some time, additional fixed points emerge because of the changed response dynamics.
In terms of dynamical systems, the method allows the forecasting of emerging bifurcations leading to new fixed points.
The additional fixed points display exactly the affected wind speed and the qualitative change in the power output.  
We have presented examples of increased and decreased power output cases due to a failure of the control system, which should limit the power output to the rated power or which should keep the turbine in an optimal operating state.
In comparison to the IEC power curve, we have given evidence that the Langevin approach detects the deviations earlier and more specifically. 
The use of high frequency noisy data will result in an information gain.
The locality of the Langevin method in phase space ($u$-$P$ space) will allow to pin down anomalies to their corresponding  velocity and power values, which is helpful for the search of the underlying cause.
Important to note is that it is also possible to extend the Langevin analysis by conditioning on further quantities.

The accuracy of the LPC is a local quantity and may change if faults emerge. 
As we have shown, the accuracy can be estimated by the data analysis itself. 
Thus, it may become necessary to adapt sampling rate and/or the amount of collected data to achieve a desired precision.

This paper shows that the Langevin approach represents a powerful tool with a broad variety of applications. 
It can be used to quantify and validate also numerical wind turbine models and characterize their dynamics. 
Because of the high sensibility of the LPC to changes in the response dynamics, the Langevin approach is usable for the characterization and to support the condition monitoring of wind turbines. 
The presented results show evidence that the LPC should also be able to detect systematic errors, e.g. in the measurement set-up.
However, in order to apply our results from numerical modelling on real data, more work has to be performed regarding the sensitivity of the LPC.
It is essential to distinguish between the uncertainties of the fixed points and small deviations due to failures in the system.
This should be a topic of further research.\\

%

{\bf Acknowledgements} The authors acknowledge helpful discussions with Julia Gottschall and thank the land of Lower Saxony and the German Ministry for Environment for the funding of this research project. For the preparation of this manuscript JP thanks for the hospitality of the Centro Internacional de Ciencias (CIC) in Cuernavaca, Mexico during the workshop ''Gone With the Wind''.

\bibliographystyle{wileyj}
\bibliography{Literatur_pc}

\end{document}